\documentclass[journal=macromolecules,manuscript=article]{achemso}

\usepackage{chemformula} % Formula subscripts using \ch{}
\usepackage[T1]{fontenc} % Use modern font encodings
\usepackage{amsmath, bm}
\author{Luofu Liu}
\affiliation{Department of Chemical and Biomolecular Engineering, University of California Berkeley, Berkeley, California 94720, United States}

\author{Rui Wang}
\email {ruiwang325@berkeley.edu}
\affiliation{Department of Chemical and Biomolecular Engineering, University of California Berkeley, Berkeley, California 94720, United States}
\alsoaffiliation{Materials Sciences Division, Lawrence Berkeley National Lab, Berkeley, California 94720, United States}

\title[]{A Molecular Theory for Liquid Crystal Elastomers: Nematic Ordering, Shape Deformation and Mechanical Response}

\begin{document}


\begin{tocentry}

\includegraphics[width=1.05\textwidth]{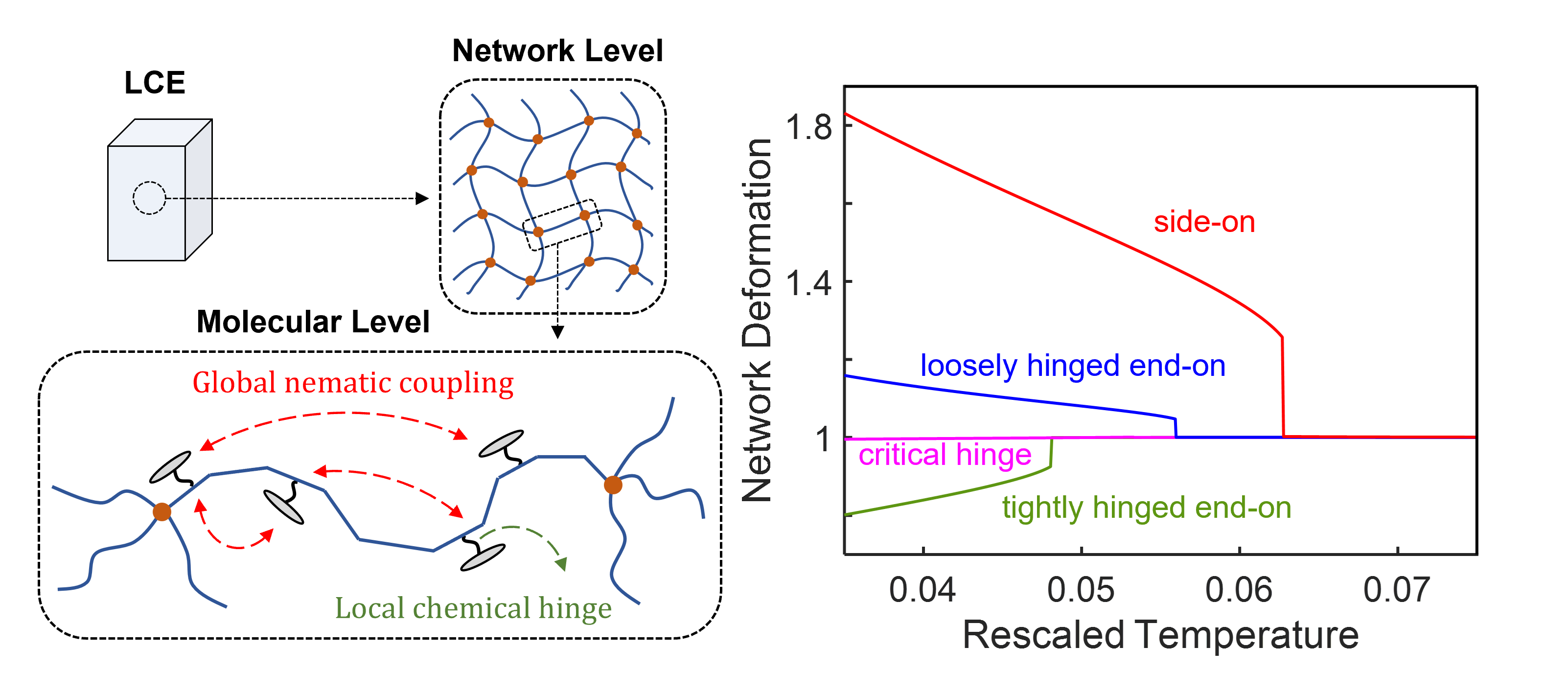}

\end{tocentry}

\begin{abstract}
Modeling liquid crystal elastomers (LCEs) at the molecular level is crucial for the predictable design of energy-conversion and stimuli-responsive materials. Here, we develop a self-consistent field theory for LCEs which captures the coupling between nematic ordering, backbone alignment and network deformation. Molecular features such as density of elastic strands, strength and architecture of local chemical hinge, and LC grafting density are systematically included. Crosslinking suppresses nematic ordering as a result of the elastic energy stored during network deformation. Higher work capacity can be achieved by less crosslinked LCEs. The spontaneous shape change of end-on side-chain LCEs can be either elongation or contraction depending on the competition between the local and global couplings. Adjusting LC grafting density is found to be an effective way to fine-tune the deformation mode. We elucidate a universal scaling relationship between the transition temperature and the shear modulus as $T_{\rm NI,0}-T_{\rm NI}\sim {\mu^*}^{\frac{4}{5}}$. Furthermore, we predict that the first-order nematic phase transition can be degraded into a continuous manner upon applied stress. Coupled with nematic ordering, the mechanical response of LCEs significantly deviates from classical rubber elasticity. A plateau in the stress-deformation curve appears accompanied by the nematic phase transition. Our theoretical predictions are in good agreement with the experimental results reported in the literature.
\end{abstract}


\section{Introduction}
There is an ever-increasing societal demand for efficient energy storage and conversion. Liquid crystal elastomers (LCEs) are crosslinked liquid crystal polymers (LCPs), where spontaneous shape change can be triggered by the ordering of LC moieties \cite{warner2007}. This ordering forces the polymer backbone into a stretched/compressed conformation which relaxes into the random coil when the ordering disappears. Such reversible shape change not only enables the energy conversion from thermal/optical/electrical/chemical forms to elastic energy, but also makes LCEs promising stimuli-responsive smart materials \cite{Stuart2010} in the applications like artificial muscles \cite{deGennes1997, Li2006, Tian2018, Chen2020}, actuators \cite{Ohm2010, Kularatne2017, Ko2017, Guin2018}, sensors \cite{Ohm2010, Shafiq2020, Mistry2020}, micro-pumps \cite{Fleischmann2012}, and adhesives \cite{Ko2017, Guo2023}, etc.

The phase and mechanical behaviors of LCEs strongly depend on their molecular structure and interactions. Compared to the uncrosslinked LCPs, it has been observed that crosslinking suppresses the nematic ordering by lowering the transition temperature and reducing the order parameter \cite{Disch1994}. The first-order feature of the phase transition can even degrade to a continuous one \cite{Tajbakhsh2001, Disch1994, Cordoyiannis2007, Cordoyiannis2009}. Recent experiments show a significant dependence of the transition temperature, actuation strain, and work capacity on the molecular details of the LCEs, including the LC content, crosslinker functionality and crosslinking density \cite{Donato2023,Barnes2022, Saed2017, Saed2017soft, Saed2019}. While mesogens can be incorporated into the polymer backbone as main-chain LCEs, connecting mesogens to the backbone as a pendant group through a flexible spacer (side-chain LCEs) facilitates the versatile design of materials: the network backbone, mesogen, and spacer can be chosen separately \cite{Rogez2018, Herbert2022}. It is interesting that the deformation mode of side-chain LCEs (SCLCEs) is highly related to the local architecture of the attachment. The spontaneous deformation of side-on SCLCEs is always elongation, in contrast, that of end-on SCLCEs can be either elongation or contraction \cite{Greve2001, Davis1993, Xu2021}. The response of LCEs to applied stress is more complicated. It depends on the geometric relationship between the force direction and the nematic director \cite{Schatzle1989, Bladon1993, Mihai2021, Warner1994, Ware2016}.

Modeling LCEs at the molecular level is crucial for the predictable design of LCE-based materials. The pioneering theory to describe the phase and mechanical behaviors of LCEs is developed by Warner, Terentjev and co-workers, \cite{Schatzle1989, Warner1996, Finkelmann2001, Warner1991, Warner1988, Zhang2019, Zhao2023} which combines the nematic energy in the Landau-de Gennes form \cite{de1993} and the neo-classical elastic energy. Such a combination provides an initial attempt to model the coupling between the nematic ordering and network deformation. However, it relies on phenomenological parameters that lack clear physical significance in the Landau energy expansion, limiting its capability to account for molecular features. An alternative approach is to replace the Landau-de Gennes form of nematic energy with the Maier-Saupe form.\cite{Liarte2011, Warsono2015, Pasini2005, Corbett2006, Corbett2008, Bai2020, Matsuyama2001, Urayama2005, Cheewaruangroj2015} This enables the applications to systems that have multiple nematic order parameters, e.g., LCEs consisting of mixture of mesogens and networks swelling in nematic solvents.\cite{Matsuyama2001, Urayama2005, Cheewaruangroj2015} The Maier-Saupe form also facilitates the description of the response under various external stimuli, such as light \cite{Corbett2006, Corbett2008, Bai2020}, electric and magnetic fields \cite{Warsono2015, Matsuyama2001}. However, all the existing theories fail to distinguish the alignment of polymer segments from mesogen orientation, which greatly hinders their application to SCLCEs. This problem is particularly severe for the end-on architecture where the local geometry prefers a perpendicular alignment between mesogens and the backbone. Furthermore, it is difficult for the existing theories to model complex LCEs with spatial inhomogeneity due to local defects or microdomains.\cite{XGWang2016} In a previous work, Wang and Wang developed a molecular model to describe the global and local coupling effects on the phase behavior and chain conformation in side-chain liquid crystal polymers.\cite{Wang2010} However, how to systematically incorporate this model into the crosslinked polymers and properly account for network elasticity remains a great challenge.

In this work, we develop a self-consistent field theory for LCEs, which is built upon the previous work by Wang and Wang on liquid crystal polymers.\cite{Wang2010} The theory systematically captures the coupling between nematic ordering, backbone alignment, and network deformation. The couplings between backbone and mesogen are treated separately by a global term associated with nematic field, and a local term originating from chemical linkage. The structure information at the molecular level, including crosslinking density, mesogen grafting density, as well as the geometry and strength of mesogen attachment, are explicitly incorporated. The phase behavior and shape deformation both in the spontaneous case and under external stress are investigated in a unified framework.

\section{Theory}

\begin{figure}[t] 
\centering
\includegraphics[width=0.6\textwidth]{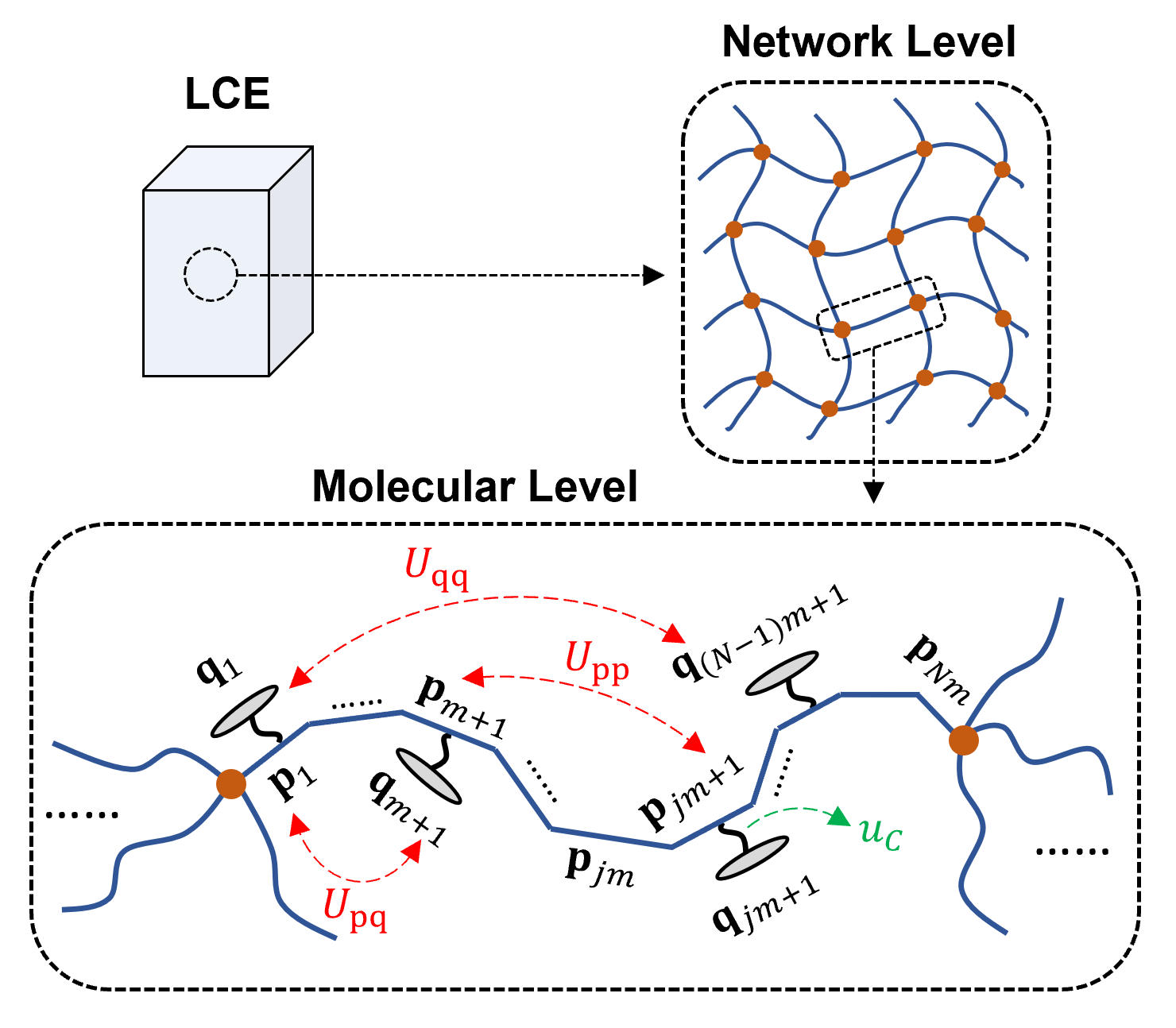}
\caption{Schematic of modeling liquid crystal elastomers (LCEs). The elastomer is a polymer network described by the neo-classical model. Each elastic strand between two crosslinks is modeled as a freely-jointed chain. Global nematic interactions between mesogens ($\bm{q}$) and backbone segments ($\bm{p}$) are denoted by red arrows. The chemical hinge connecting the mesogen and its attached backbone segment is considered as the local coupling (green arrows).} 
%\vspace{-2em}
\label{figure:schematic}
\end{figure}

As illustrated in Fig. \ref{figure:schematic}, the LCE is modeled by a network formed by end-linking $n$ liquid crystal polymers. Here, we consider a general case of SCLCEs. Each LCP between two adjacent crosslinkers is described by the freely-jointed chain model with $N$ repeating sections. Each section contains $m$ polymer backbone segments and one mesogen, with a Kuhn length $b$. The number of segments for one entire LCP is thus $Nm$. Without loss of generality, the mesogen is assumed to be grafted on the first segment of the repeating section. $m$ hence represents the mesogen grafting density where $m=1$ is the fully grafted case. We use $\bm{p}_{\alpha, 1}$, $\bm{p}_{\alpha, 2}$, ..., $\bm{p}_{\alpha, Nm}$ and $\bm{q}_{\alpha, 1}$, $\bm{q}_{\alpha, 2}$, ..., $\bm{q}_{\alpha, (N-1)m+1}$ to denote the unit orientational vectors of the backbone segment and mesogen in the $\alpha$-th chain. The model recovers MCLCE in the absence of the grafted mesogens (i.e., all the $\bm{q}$ vectors).

We start by defining the following microscopic operators to describe the local density and molecular orientation of the backbone segment and mesogen. The density operators are given by
\begin{equation}\label{eq:density_operators}
\begin{aligned}
    &\hat{\phi}_{\rm p}(\bm{r}) = v_{\rm{p}}\sum_{\alpha=1}^{n}\sum_{j=1}^N\delta(\bm{r} - \bm{r}_{\alpha,(j-1)m+1})\\
    &\hat{\phi}_{\rm p^*}(\bm{r}) = v_{\rm{p}}\sum_{\alpha=1}^{n}\sum_{j=1}^N\sum_{k=2}^m\delta(\bm{r} - \bm{r}_{\alpha,(j-1)m+k})\\
    &\hat{\phi}_{\rm q}(\bm{r}) = v_{\rm{q}}\sum_{\alpha=1}^{n}\sum_{j=1}^N\delta(\bm{r} - \bm{r}_{\alpha,(j-1)m+1})
\end{aligned}
\end{equation}
where $\rm p$, $\rm p^*$, $\rm q$ refer to the LC-grafted backbone segment, non-grafted segment, and mesogen, respectively. $v_{\rm p}$ and $v_{\rm q}$ are the occupied volume of backbone segment and mesogen. it should be noted that the volume of the spacer connecting the mesogen with the backbone is not accounted for in this work. As discussed later, the spacer will be introduced as a local hinge effect. The position vector of the $k$-th segment in the $j$-th repeating section on the $\alpha$-th chain is given by $\bm{r}_{\alpha,(j-1)m+k}=\bm{R}_{\alpha,0} + b\sum_{k=1}^{(j-1)m+k-1}\bm{p}_{\alpha,k}+(b/2)\bm{p}_{\alpha,(j-1)m+k}$ with $\bm{R}_{\alpha,0}$ the position of one of the chain-end. The nematic operators are given by
\begin{equation}\label{eq:orderparam_operators}
\begin{aligned}
    &\hat{\bm{S}}_{\rm p}(\bm{r}) = v_{\rm{p}}\sum_{\alpha=1}^{n}\sum_{j=1}^N\delta(\bm{r} - \bm{r}_{\alpha,(j-1)m+1})\left[\bm{p}_{\alpha,(j-1)m+1}\otimes\bm{p}_{\alpha,(j-1)m+1}-\frac{1}{3}\bm{I}\right]\\
    &\hat{\bm{S}}_{\rm p^*}(\bm{r}) = v_{\rm{p}}\sum_{\alpha=1}^{n}\sum_{j=1}^N\sum_{k=2}^m\delta(\bm{r} - \bm{r}_{\alpha,(j-1)m+k})\left[\bm{p}_{\alpha,(j-1)m+k}\otimes\bm{p}_{\alpha,(j-1)m+k}-\frac{1}{3}\bm{I}\right]\\
    &\hat{\bm{S}}_{\rm q}(\bm{r}) = v_{\rm{q}}\sum_{\alpha=1}^{n}\sum_{j=1}^N\delta(\bm{r} - \bm{r}_{\alpha,(j-1)m+1})\left[\bm{q}_{\alpha,(j-1)m+1}\otimes\bm{q}_{\alpha,(j-1)m+1}-\frac{1}{3}\bm{I}\right]
\end{aligned}
\end{equation}
where $\bm{I}$ stands for the identity tensor. 

Following the treatment by Wang and Wang, the coupling between the chain backbone and LC side groups is properly accounted for by two types of terms: the global nematic interaction and the local chemical hinge effect.\cite{Wang2010}  The Hamiltonian of the system consists of nematic, hinge, and elastic contributions:
\begin{equation}
\mathcal{H} = \mathcal{H}_{\rm nem} + \mathcal{H}_{\rm hinge} + \mathcal{H}_{\rm ela}
\end{equation}
The nematic interaction is described by two-body pseudopotentials between selected pairs of mesogen, grafted backbone segment and non-grafted backbone segment in the Maier-Saupe type,\cite{Wang2010,Zhuang2012,Maier1959} which yields
\begin{equation} \label{eq:nematic_Hamiltonian}
\begin{aligned}
\mathcal{H}_{\rm nem} &= -\frac{U_{\rm pp}}{2}\int {\rm d}\bm{r}\hat{\bm{S}}_{\rm p}(\bm{r}):\hat{\bm{S}}_{\rm p}(\bm{r}) -\frac{U_{\rm p^{*}p^{*}}}{2}\int {\rm d}\bm{r}\hat{\bm{S}}_{\rm p^*}(\bm{r}):\hat{\bm{S}}_{\rm p^*}(\bm{r})  -\frac{U_{\rm qq}}{2}\int {\rm d}\bm{r}\hat{\bm{S}}_{\rm q}(\bm{r}):\hat{\bm{S}}_{\rm q}(\bm{r})\\
&-U_{\rm pp^*}\int {\rm d}\bm{r}\hat{\bm{S}}_{\rm p}(\bm{r}):\hat{\bm{S}}_{\rm p^*}(\bm{r}) -U_{\rm pq}\int {\rm d}\bm{r}\hat{\bm{S}}_{\rm p}(\bm{r}):\hat{\bm{S}}_{\rm q}(\bm{r}) -U_{\rm p^*q}\int {\rm d}\bm{r}\hat{\bm{S}}_{\rm p^*}(\bm{r}):\hat{\bm{S}}_{\rm q}(\bm{r})
\end{aligned}
\end{equation}
where $U_{\kappa \kappa'}$ is the nematic interaction parameter between species $\kappa$ and $\kappa'$ ($\kappa,\kappa'={\rm p},{\rm p^*},{\rm q}$). 

The hinge effect accounts for the local coupling between the LC side group and its attached chain segment as a result of chemical connection from the spacer. For simplicity, we assume that the local coupling only exists between the mesogen and the backbone segment it directly connects to. Its effect on the adjacent backbones is neglected in this work. The hinge effect is modeled by a simple quadratic form as
\begin{equation}
 \mathcal{H}_{\rm hinge} = -\sum_{\alpha=1}^{n}\sum_{j=1}^N u_C(\bm{p}_{\alpha, (j-1)m+1}\cdot \bm{q}_{\alpha, (j-1)m+1})^2
\end{equation}
where $u_C$ is the coupling parameter. The magnitude of $u_C$ characterizes the strength of the local coupling, which is determined by factors like spacer length and stiffness. The sign of $u_C$ reflects the geometry of the alignment between the mesogen ($\bm{q}_{\alpha, (j-1)m+1}$) and the backbone ($\bm{p}_{\alpha, (j-1)m+1}$). $u_C>0$ represents the case of side-on SCLCEs where the mesogen prefers to align parallel to its attached backbone. $u_C < 0$ represents the case of end-on SCLCEs where the perpendicular alignment is energetically favorable.

We consider the LCE as a uniformly crosslinked ideal network where topological defects such as dangling ends and cyclic loops are ignored. The elastic energy is described by the neo-classical model \cite{warner2007}, given by
\begin{equation} \label{eq:elastic_Hamiltonian}
\mathcal{H}_{\rm ela} = \frac{1}{2}\mu \int {\rm d}\bm{r} {\rm Tr} [\bm{l}_0 \bm{\lambda}^{\rm T}(\bm{r}) \hat{\bm{l}}^{-1}(\bm{r}) \bm{\lambda}(\bm{r}) ]
\end{equation}
where $\mu$ is the shear modulus proportional to the density of the elastic strands. $\bm{l}_0$ and $\hat{\bm{l}}$ are the step length tensor for the as-prepared state and deformed state, respectively. $\bm{\lambda}$ is the deformation gradient tensor. $\hat{\bm{l}}$ can be related to the orientations of backbone segments as
\begin{equation}
\hat{\bm{l}}(\bm{r}) = b\left\{\bm{I} + \frac{1}{m} \left[\frac{3\hat{\bm{S}}_{\rm p}(\bm{r})}{\hat{\phi}_{\rm p}(\bm{r})} \right] + \frac{m-1}{m} \left[\frac{3\hat{\bm{S}}_{\rm p^*}(\bm{r})}{\hat{\phi}_{\rm p^*}(\bm{r})} \right] \right\}
\end{equation}
The prefactors $3/\hat{\phi}_{\kappa}$ are introduced for the consistency with the standard definition of order parameter tensors.\cite{warner2007} The terms $1/m$ and $(m-1)/m$ weight the contributions from grafted and non-grafted backbone segments to the step length based on their number fractions.

For LCEs under an applied nominal stress $\bm{\sigma}_{\rm n}$, the partition function of the system is then calculated by integrating all configurations of segments and mesogens as well as different deformations:
\begin{equation} \label{eq:total_partition_function}
\begin{aligned}
Z &= \int {\rm D}\bm{\lambda}\exp \left[
\beta \int {\rm d}\bm{r} \bm{\sigma}_{\rm n}:\bm{\lambda}(\bm{r}) \right] 
\prod_{\alpha=1}^{n}\left(
\int {\rm d}\bm{R}_{\alpha,0} \prod_{j=1}^N \prod_{k=1}^m \int {\rm d} \bm{p}_{\alpha, (j-1)m+k} \cdot \prod_{j=1}^N \int {\rm d} \bm{q}_{\alpha, (j-1)m+1} \right)\\
&\prod_{\bm{r}} \delta\left[\hat{\phi}_{\rm p}(\bm{r}) +\hat{\phi}_{\rm p^*}(\bm{r}) + \hat{\phi}_{\rm q}(\bm{r}) -1 \right] \cdot \prod_{\bm{r}}\delta \left[ \det \bm{\lambda}(\bm{r}) - 1 \right] \exp(-\beta \mathcal{H})
\end{aligned}
\end{equation}
The functional integral over $\bm{\lambda}$ results from the transform of ensemble from constant deformation to constant applied stress. The first $\delta$-functional enforces the local incompressibility, whereas the second one originates from the volume conservation of incompressible materials under deformation.

We follow the standard self-consistent field approach which involves (1) identity transformation to decouple the many-body interactions into a single chain interacting with fluctuating fields and (2) replacing the functional integral with its saddle-point value \cite{fredrickson2006equilibrium}. The detailed derivation is provided in Sec. I of Supporting Information. The theory is general for LCEs with spatial inhomogeneity in both composition and nematic ordering. For a simple case of homogeneous networks, the self-consistent field theory ultimately leads to the computation of the single-chain partition function in spatially homogeneous fields $\bm{M}_\kappa$ as
\begin{equation}\label{eq:bulk_single_chain_partition_function}
\begin{aligned}
Q &= \prod_{j=1}^N \prod_{k=1}^m \int {\rm d} \bm{p}_{ (j-1)m+k} \cdot \prod_{j=1}^N \int {\rm d} \bm{q}_{ (i-1)m+1} \exp\left[\sum_{j=1}^N \beta u_C(\bm{p}_{ (j-1)m+1} \cdot \bm{q}_{ (j-1)m+1})^2 \right] \\
&\exp \bigg\{ \sum_{j=1}^N \left[\beta v_{\rm p} \bm{M}_{\rm p} : \left(\bm{p}_{(j-1)m+1} \otimes \bm{p}_{(j-1)m+1} -\frac{1}{3} \bm{I} \right)  \right] \\
&+ \sum_{j=1}^N \sum_{k=2}^m \left[\beta v_{\rm p} \bm{M}_{\rm p^*} : \left(\bm{p}_{(j-1)m+k} \otimes \bm{p}_{(j-1)m+k} -\frac{1}{3} \bm{I} \right)  \right] \\
&+\sum_{j=1}^N \left[\beta v_{\rm q} \bm{M}_{\rm q} : \left(\bm{q}_{(j-1)m+1} \otimes \bm{q}_{(j-1)m+1} -\frac{1}{3} \bm{I} \right)  \right] \bigg\} \\
\end{aligned}
\end{equation}
where
\begin{subequations}
\begin{align}
&\bm{M}_{\rm p} =  U_{\rm pp} \bm{S}_{\rm p} + U_{\rm pp^*} \bm{S}_{\rm p^*} + U_{\rm pq} \bm{S}_{\rm q} + \frac{3\mu}{2m\phi_{\rm p}} b \bm{l}^{-1}\bm{\lambda}\bm{l}_0\bm{\lambda}^{\rm T}\bm{l}^{-1}\\
&\bm{M}_{\rm p^*} =  U_{\rm pp^*} \bm{S}_{\rm p} + U_{\rm p^*p^*} \bm{S}_{\rm p^*} + U_{\rm p^*q} \bm{S}_{\rm q} + \frac{3(m-1)\mu}{2m\phi_{\rm p^*}} b\bm{l}^{-1}\bm{\lambda}\bm{l}_0\bm{\lambda}^{\rm T}\bm{l}^{-1}\\
&\bm{M}_{\rm q}=U_{\rm pq} \bm{S}_{\rm p} + U_{\rm p^*q} \bm{S}_{\rm p^*} + U_{\rm qq} \bm{S}_{\rm q}
\end{align}
\end{subequations}
$\bm{S}_{\kappa}$ is the tensorial order parameter corresponding to the nematic operators. The equilibrium values of $\bm{S}_\kappa$ and $\bm{\lambda}$ can be obtained from the saddle-point approximation, which are given by
\begin{subequations}\label{eq:SC_Equations}
\begin{align}
&\bm{S}_{\rm p} = \frac{n v_{\rm p}}{V} \left \langle\sum_{j=1}^N\left(\bm{p}_{(j-1)m+1}\otimes\bm{p}_{(j-1)m+1} - \frac{1}{3} \bm{I} \right) \right \rangle \label{eq:Sp}\\
&\bm{S}_{\rm p^*} = \frac{n v_{\rm p}}{V} \left \langle\sum_{j=1}^N \sum_{k=2}^m\left(\bm{p}_{(j-1)m+k}\otimes\bm{p}_{(j-1)m+k} - \frac{1}{3} \bm{I} \right) \right \rangle \label{eq:Spstar}\\
&\bm{S}_{\rm q} = \frac{n v_{\rm q}}{V} \left \langle\sum_{j=1}^N\left(\bm{q}_{(j-1)m+1}\otimes\bm{q}_{(j-1)m+1} - \frac{1}{3} \bm{I} \right) \right \rangle \label{eq:Sq} \\
&\bm{l}_0 \bm{\lambda}^{\rm T}\bm{l}^{-1} + \bm{l}_0^{\rm T} \bm{\lambda}^{\rm T}\bm{l}^{-\rm T} - \gamma  \bm{\lambda}^{-1} -2\mu^{-1}\bm{\sigma}_{\rm n}= \bm{0} \label{eq:deformation}
\end{align}
\end{subequations}
where $\langle ... \rangle$ is the ensemble average evaluated based on the Boltzmann factor in the single-chain partition function in Eq. \ref{eq:bulk_single_chain_partition_function}. $\gamma$ in Eq. \ref{eq:deformation} comes from Lagrangian multiplier as a result of volume conservation. In principle, Eqs. \ref{eq:SC_Equations} can describe LCEs with the nematic ordering and applied stress in any direction. In this work, we focus on the case of uniaxial order and deformation along the $z$-axis. Both $\bm{\lambda}$ and $\bm{\sigma}_{\rm n}$ are thus diagonal: $\lambda_{zz} = \lambda$, $\lambda_{xx} = \lambda_{yy}=1/\sqrt{\lambda}$, and $\sigma_{{\rm n},zz}=\sigma_{\rm n}$, $\sigma_{{\rm n},xx}=\sigma_{{\rm n},yy}=0$. The tensorial order parameters are diagonal and traceless, with their scaler forms defined by normalizing the $zz$-component by the volume fractions:
\begin{subequations}\label{eq:SC_Equations_scaler}
\begin{align}
&s_{\rm p} = \frac{3}{2}\frac{\bm{S}_{{\rm p},zz}}{\phi_{\rm p}} =\frac{1}{N} \left \langle\sum_{j=1}^N P_2 \left(\cos \theta_{{\rm p},(j-1)m+1} \right) \right \rangle \label{eq:sp}\\
&s_{\rm p^*} = \frac{3}{2}\frac{\bm{S}_{{\rm p^*},zz}}{\phi_{\rm p^*}} =\frac{1}{N(m-1)} \left \langle\sum_{j=1}^N \sum_{k=2}^m P_2 \left(\cos \theta_{{\rm p^*},(j-1)m+k} \right) \right \rangle \label{eq:spstar}\\
&s_{\rm q} = \frac{3}{2}\frac{\bm{S}_{{\rm q},zz}}{\phi_{\rm q}} =\frac{1}{N} \left \langle\sum_{j=1}^N P_2 \left(\cos \theta_{{\rm q},(j-1)m+1} \right) \right \rangle \label{eq:sq}
\end{align}
\end{subequations}
where $P_2(x) = 3x^2/2-1/2$ is the second Legendre polynomial, and $\theta_{\kappa,i}$ is the angle between the $i$-th segment (of species $\kappa$) and the nematic director. The scaler form of Eq. \ref{eq:deformation} in uniaxial deformation becomes
\begin{equation}\label{eq:deformation_scaler}
    \sigma_{\rm n} = \mu \left( \lambda \frac{1+2\bar{s}_{\rm p,0}}{1+2\bar{s}_{\rm p}} - \frac{1}{\lambda^2} \frac{1-\bar{s}_{\rm p,0}}{1-\bar{s}_{\rm p}}\right)
\end{equation}
where $\bar{s}_{\rm p} = (1/m) s_{\rm p} + (1-1/m) s_{\rm p^*}$ is the average nematic order parameter of the backbone segments, and $\bar{s}_{\rm p,0}$ is that at the as-prepared state. It should be noted that Eq. \ref{eq:deformation_scaler} recovers the classical stress-deformation relationship for an isotropic rubber, $\sigma_{\rm n} = \mu (\lambda - 1/\lambda^2)$, if nematic ordering is absent ($\bar{s}_{\rm p} = \bar{s}_{\rm p,0} = 0$). Furthermore, Eq. \ref{eq:deformation_scaler} reduces to $\lambda^3 = [(1-\bar{s}_{\rm p,0})(1+2\bar{s}_{\rm p})]/[(1+2\bar{s}_{\rm p,0})(1-\bar{s}_{\rm p})] $ for the force-free case, which is consistent with the equation for spontaneous deformation used in the literature \cite{warner2007,Warner1988,Warner1991}.

The freely-jointed chain model together with the self-coupling form of the hinge effect allows a further decoupling of the single-chain partition function $Q$ down to the level of individual segments: $Q = (Q_1 Q_2^{m-1})^N$. $Q_1$ and $Q_2$ are the partition functions of the LC-grafted segment and the non-grafted segment respectively, which are given by
\begin{equation}\label{eq:Q1}
\begin{aligned}
Q_1 &= \int {\rm d}\bm{p} \int {\rm d}\bm{q} \exp\left[ \beta u_C(\bm{p}\cdot\bm{q})^2 \right] \\
&\exp \left \{ \frac{m+r}{1+r} \left[  \beta u_{\rm pp}\phi_{\rm p}^2 s_{\rm p} + \beta u_{\rm pp^*}\phi_{\rm p} \phi_{\rm p^*}s_{\rm p^*}  + \beta u_{\rm pq}\phi_{\rm p} \phi_{\rm q}s_{\rm q} + E  \right] P_2(\cos \theta_{\rm p})\right\}\\
&\exp \left \{\frac{m+r}{1+r} \left[\beta u_{\rm pq}\phi_{\rm p}\phi_{\rm q} s_{\rm p} + \beta u_{\rm p^*q}\phi_{\rm p^*} \phi_{\rm q}s_{\rm p^*}  + \beta u_{\rm qq}\phi_{\rm q}^2 s_{\rm q} \right] P_2(\cos \theta_{\rm q})\right\}
\end{aligned}
\end{equation}
and
\begin{equation}\label{eq:Q2}
Q_2 = \int {\rm d} \bm{p}^* \exp \left\{ \frac{m+r}{1+r} \left[ \frac{\beta u_{\rm pp^*}\phi_{\rm p}\phi_{\rm p^*} s_{\rm p} + \beta u_{\rm p^*p^*}\phi_{\rm p^*}^2 s_{\rm p^*}  + \beta u_{\rm p^*q}\phi_{\rm p^*} \phi_{\rm q} s_{\rm q}}{m-1} + E \right]P_2(\cos \theta_{\rm p^*})\right\}
\end{equation}
$E$ in Eqs. \ref{eq:Q1} and \ref{eq:Q2} is the elastic contribution:
\begin{equation}
E = \frac{\mu^*}{m} \left(\frac{\lambda^2(1+2\bar{s}_{\rm p,0})}{(1+2\bar{s}_{\rm p})^2} - \frac{1-\bar{s}_{\rm p,0}}{\lambda(1-\bar{s}_{\rm p})^2} \right)
\end{equation}
with $r=v_{\rm q}/v_{\rm p}$ the volume ratio between mesogen and backbone segment. $u_{\kappa \kappa'}=(2/3)U_{\kappa \kappa'}(v_{\rm p} + v_{\rm q})$ is the rescaled nematic interaction parameter. $\mu^* = \beta \mu (v_{\rm p} + v_{\rm q})$ is the dimensionless shear modulus which is only proportional to the density of elastic strands but independent of temperature. The elastic effect is systematically incorporated in the single-segment partition functions and thus implicitly alters the chain conformation and nematic ordering through Eq. \ref{eq:SC_Equations_scaler}. Furthermore, the three volume fractions can be expressed by two structure parameters $m$ and $r$ as $\phi_{\rm p} = 1/(m+r)$, $\phi_{\rm q} = r/(m+r)$ and $\phi_{\rm p^*}=(m-1)/(m+r)$. It is worth noting that Eqs. \ref{eq:Q1} and \ref{eq:Q2} present the molecular nature of our theory: partition functions are determined by molecular interaction parameters ($u_{\kappa \kappa'}$ and $u_C$) as well as molecular structural parameters ($m$, $r$ and $\mu^*$).

The order parameters in Eq. \ref{eq:SC_Equations_scaler} can also be evaluated at the segment level as
\begin{equation} \label{eq:SC_Equations_scaler_simplest}
    s_{\kappa} = \langle P_2(\cos \theta_\kappa) \rangle, \ \ \kappa = {\rm p,p^*,q}
\end{equation}
where the average for ${\rm p}$ and ${\rm q}$ is taken based on $Q_1$, and that for ${\rm p^*}$ is based on $Q_2$. By solving the self-consistent equations \ref{eq:deformation_scaler} and \ref{eq:SC_Equations_scaler_simplest} iteratively, we can obtain the equilibrium value of the nematic ordering, segment alignment, and network deformation under a given applied uniaxial stress (see Sec. II of Supporting Information for numerical details). The free energy of one repeating section is given by
\begin{equation}\label{eq:section_free_energy}
\begin{aligned}
\beta F_{\rm S} &= \frac{m+r}{1+r} \bigg(\frac{1}{2}\beta u_{\rm pp} \phi_{\rm p}^2 s_{\rm p}^2 +  \frac{1}{2}\beta u_{\rm p^*p^*} \phi_{\rm p^*}^2 s_{\rm p^*}^2 + \frac{1}{2}\beta u_{\rm qq} \phi_{\rm q}^2 s_{\rm q}^2 + \beta u_{\rm p p^*}\phi_{\rm p}\phi_{\rm p^*}s_{\rm p}s_{\rm p^*} \\
& + \beta u_{\rm pq} \phi_{\rm p} \phi_{\rm q}s_{\rm p} s_{\rm q} + \beta u_{\rm p^*q}\phi_{\rm p^*}\phi_{\rm q}s_{\rm p^*} s_{\rm q}\bigg) 
 - \ln Q_1 - (m-1)\ln Q_2\\
 &+ \frac{m+r}{1+r}\frac{\mu^*}{2} \left[ \lambda^2(1+2\bar{s}_{\rm p,0}) \frac{1+4\bar{s}_{\rm p}}{(1+2\bar{s}_{\rm p})^2} + \frac{2}{\lambda}(1-\bar{s}_{\rm p,0}) \frac{1-2\bar{s}_{\rm p}}{(1-\bar{s}_{\rm p})^2} - \frac{2\sigma_{\rm n}}{\mu}\lambda \right] 
\end{aligned}
\end{equation}
It should be noted that Eqs. \ref{eq:deformation_scaler} and \ref{eq:SC_Equations_scaler_simplest} can also obtained via minimization of $F_{\rm S}$ with respect to $s_{\kappa}$ and $\lambda$. 

Although the equations are presented in detail for ideal homogeneous networks under uniaxial stress along the director, our self-consistent field theory is universal for LCEs with spatial inhomogeneity and for any form of stress. The theory can easily incorporate topological defects in real networks like dangling ends and cyclic loops.\cite{rubinstein2003,Wang2016,Zhong2016} The theory can also be generalized to the swelling of LCEs in either isotropic or nematic solvents,\cite{Urayama2005isotropic, Matsuyama2001, Urayama2005, Cheewaruangroj2015} and the response of LCEs to other external stimuli such as optical, electric and magnetic.\cite{Corbett2006, Corbett2008, Bai2020,Warsono2015, Matsuyama2001}

\section{Results and Discussion}

As discussed in the above section, the macroscopically measurable properties, nematic ordering ($s_{\rm q}$) and network deformation ($\lambda$) under stress $\sigma_{\rm n}$ can be calculated based on the molecular information of interactions ($u_{\kappa \kappa'}$ and $u_C$) and structures ($m$, $r$ and $\mu^*$). The LCE theory recovers the main-chain case by ignoring the mesogen terms, i.e., setting $r=0$ and $u_C=0$. The case of spontaneous deformation is naturally recovered when  $\sigma_{\rm n}=0$. For simplicity, we here consider networks prepared at the isotropic state such that $s_{\rm p,0} = 0$.

\subsection{Effects of the Network Structure on Nematic Ordering and Spontaneous Deformation}

\begin{figure}[h] 
\centering
\includegraphics[width=1\textwidth]{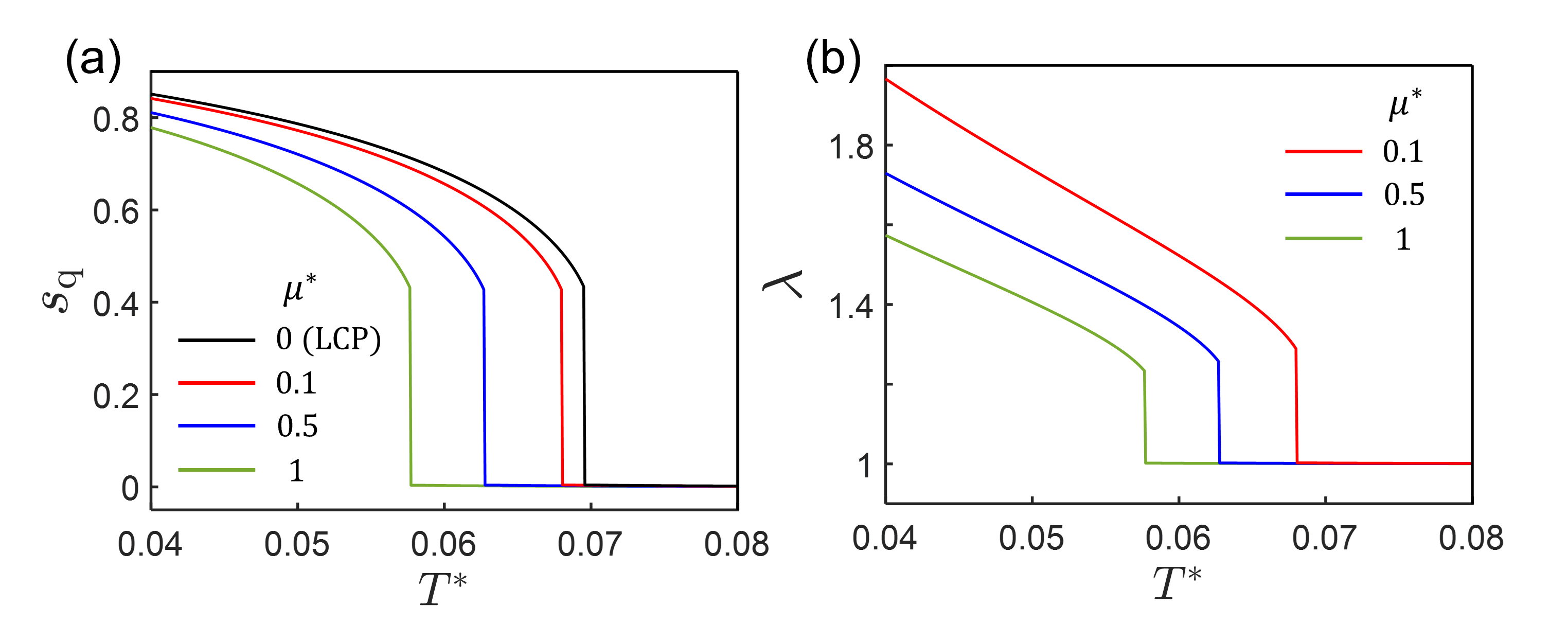}
\caption{Effect of density of elastic strands on the mesogen ordering and spontaneous shape change under the force-free condition ($\sigma_{\rm n}=0$). (a) The nematic order parameter of mesogen $s_{\rm q}$ and (b) network deformation $\lambda$ are plotted as a function of rescaled temperature for various dimensionless shear moduli $\mu^*$. The temperature is in a rescaled form as $T^* = k_{\rm B}T/u_{\rm qq}$. $m=1$, $r=1$, $u_{\rm pp}=0.07u_{\rm qq}$, $u_{\rm pq}=0.2u_{\rm qq}$, and $u_C = 0.25u_{\rm qq}$.} 
\label{figure:figure2}
\end{figure}

In this subsection, we focus on the interplay between the nematic ordering and spontaneous shape change in the absence of applied stress. The molecular nature of our theory facilitates the systematic examination of different aspects of the network structure, including the density of elastic strands, the strength and architecture of chemical hinge, and the LC grafting density. Figure \ref{figure:figure2} plots the nematic order parameter of mesogen $s_{\rm q}$ and network deformation $\lambda$ as a function of temperature for different densities of elastic strands in side-on SCLCEs. The density of elastic strands is represented by the dimensionless shear modulus $\mu^*$, where $\mu^*=0$ stands for the special case of uncrosslinked LCP. As $\mu^*$ increases, Fig. \ref{figure:figure2}a shows that the nematic-isotropic transition shifts to a lower temperature, while the order parameter in the nematic phase is also reduced. This trend predicted by the theory is consistent with the experimental observation by Disch \textit {et al}.\cite{Disch1994} The suppression of nematic ordering by crosslinking originates from the requirement to store extra energy in the elastic form. For SCLCEs, this suppression is realized through the coupling between the backbone and mesogen, by means of both the global nematic field and the local chemical hinge. The coupling effect critical for SCLCEs cannot be captured by existing theories.

The isotropic-to-nematic transition is accompanied by the spontaneous shape change of the elastomer as shown in Fig. \ref{figure:figure2}b. The deformation becomes smaller as $\mu^*$ increases, because the network with a higher modulus has a larger resistance to deform. Our results suggest that less crosslinked LCE is preferred to achieve higher work capacity after returning to its isotropic state due to larger spontaneous shape change. The effect of density of elastic strands on the LCE work capacity predicted by our theory is also in good agreement with the experimental findings by Saed \textit{et al}.\cite{Saed2017}

\begin{figure}[h] 
\centering
\includegraphics[width=1\textwidth]{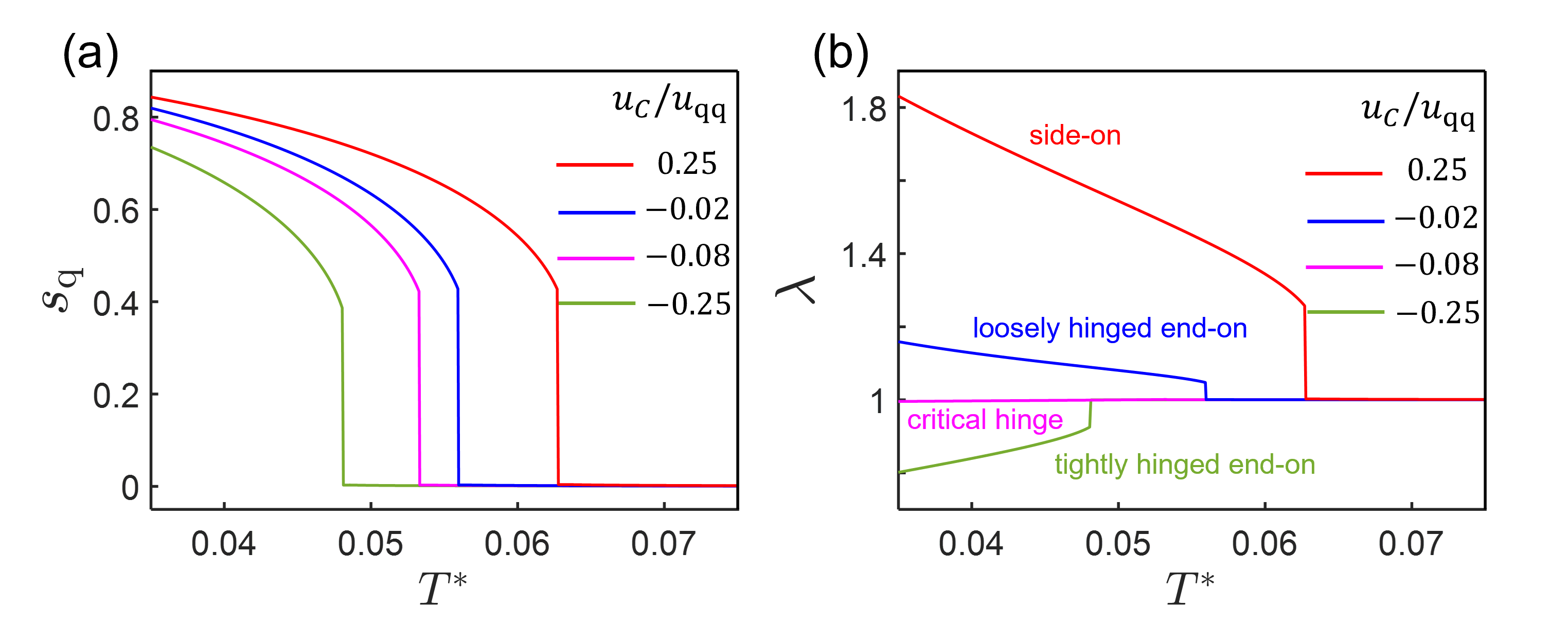}
\caption{Effect of the local chemical hinge on the mesogen ordering and spontaneous shape change. (a) $s_{\rm q}$ and (b) $\lambda$ are plotted as a function of $T^*$ for different values of local coupling parameter $u_C$. $m=1$, $r=1$, $u_{\rm pp}=0.07u_{\rm qq}$, $u_{\rm pq}=0.2u_{\rm qq}$, and $\mu^*=0.5$.} 
\label{figure:figure3}
\end{figure}

It has been observed in experiments that how the mesogen is attached to the backbone in LCEs has a profound effect on both the phase behavior and shape change.\cite{Greve2001,Xu2021} While the spontaneous shape change of side-on SCLCEs is always elongation, that of the end-on SCLCEs can be either elongation or contraction depending on the chemical hinge and other structural parameters. The chemical hinge effect is modeled by the local coupling term in our theory, where the coupling parameter $u_C$ characterizes the geometry and the strength of the hinge. $u_C>0$ stands for the side-on attachment whereas $u_C<0$ stands for the end-on case. Figure \ref{figure:figure3} depicts the local coupling effect on the mesogen ordering and network deformation. As shown in Fig. \ref{figure:figure3}a, the nematic ordering of mesogen is weakened with a corresponding reduction of the transition temperature, when the hinge geometry changes from side-on to end-on. This is because the preferable perpendicular alignment of the mesogen and backbone in the end-on geometry leads to a disturbance to the global nematic field. Furthermore, Fig. \ref{figure:figure3}b shows that the elongational shape change of the side-on SCLCE also becomes less pronounced as $u_C$ decreases and eventually turns to contraction (manifested by $\lambda < 1$) when $u_C$ gets more negative. At $u_C=-0.08u_{\rm qq}$, it is interesting to note that the end-on SCLCE does not exhibit any spontaneous shape change in the entire temperature range even though nematic ordering occurs at low temperatures. For this critical $u_C$, the local hinge effect exactly offsets the global nematic field, which leads to effective decoupling of the backbone segments from mesogens.

The underlying mechanism that end-on SCLCEs can undergo either elongation or contraction is a result of the competition between the global and local couplings. While the global nematic coupling prefers the parallel alignment of both mesogen and backbone along the director, the local coupling restrained by the chemical hinge favors a perpendicular alignment. For a small negative value of $u_C$ (loosely hinged end-on), the global coupling dominates, leading to elongation. On the other hand, the local coupling becomes more important as $u_C$ gets more negative (tightly hinged end-on), which enforces a contractive deformation. Our theory can well capture the trend found by Xu \textit{et al.} that the spontaneous elongation of LCEs changes to contraction by gradually replacing side-on mesogens with end-on ones.\cite{Xu2021} Our theory can also explain the fact that LCE with the same end-on mesogen but different network structures can undergo either elongation or contraction.\cite{Greve2001}

\begin{figure}[h] 
\centering
\includegraphics[width=0.7\textwidth]{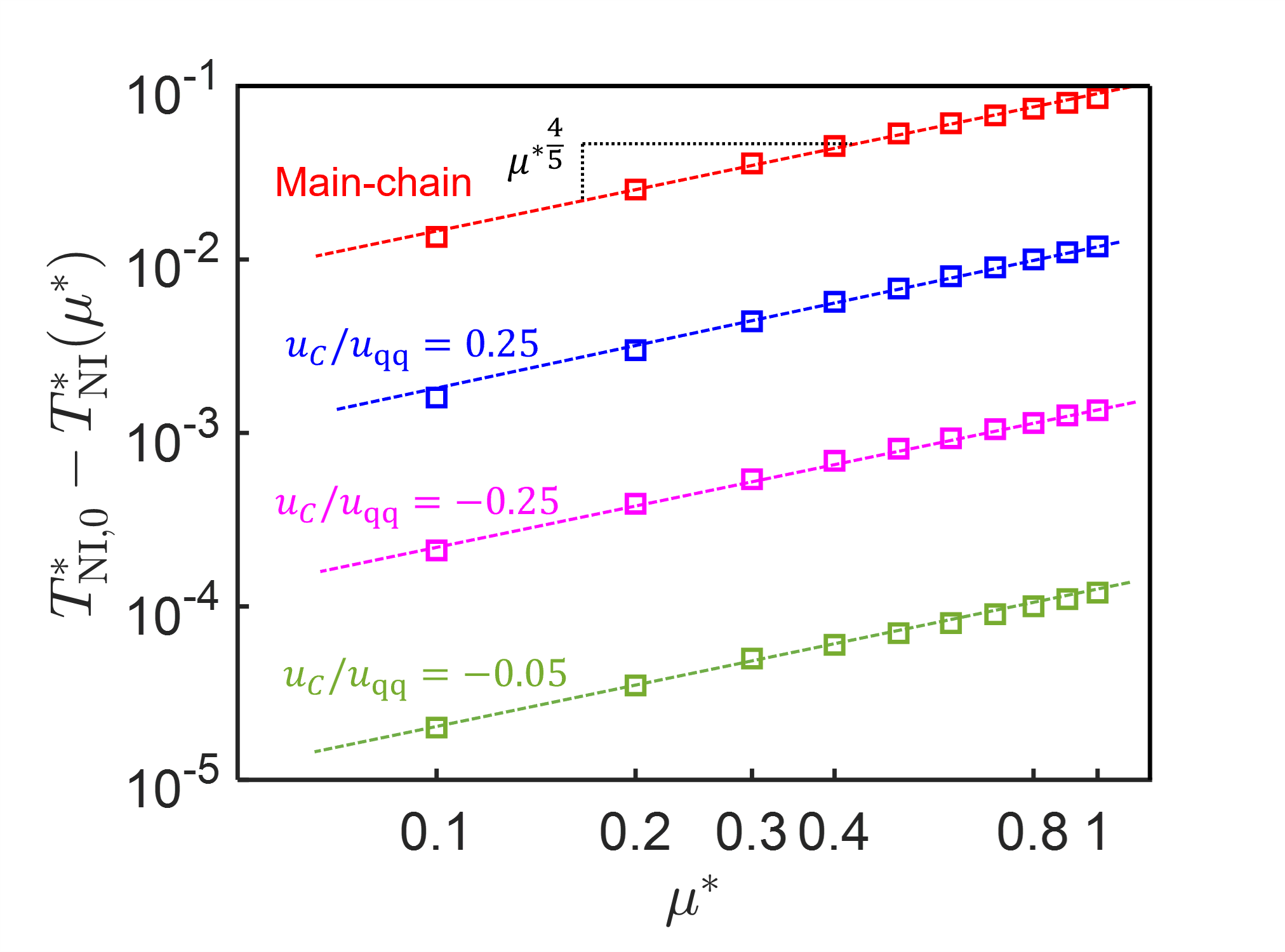}
\caption{The reduction of nematic-isotropic transition temperature as a function of dimensionless shear modulus for MCLCE and SCLCEs with different hinge effects. $T_{\rm NI,0}^*$ is the rescaled transition temperature for uncrosslinked LCPs. The results are presented in the log-log plot, leading to a universal exponent $4/5$. For SCLCEs, $m=1$, $r=1$, $u_{\rm pp}=0.07u_{\rm qq}$, $u_{\rm pq}=0.2u_{\rm qq}$.} 
\label{figure:figure4}
\end{figure}

Figure \ref{figure:figure4} summarizes the crosslinking-induced reduction of the transition temperature $T_{\rm NI}$ from that of the uncrosslinked LCP ($T_{\rm NI,0}$). The results of both MCLCE and SCLCEs with different hinge effects are presented. The reduction of $T_{\rm NI}$ is most significant in MCLCE because mesogens are directly incorporated into the backbone and hence have the largest impact on network deformation. For SCLCEs, The reduction of $T_{\rm NI}$ shows a non-monotonic dependence on the local coupling parameter $u_C$ due to the competition between the global and local coupling effects. In the limiting case of loosely hinged end-on SCLCE (e.g. $u_C/u_{\rm qq}=-0.05$), the two coupling effects almost cancel each other, resulting in comparably small deformation (see Fig. \ref{figure:figure3}b) and negligible reduction of $T_{\rm NI}$. Although the reduction of $T_{\rm NI}$ is different for different LCE structures, it follows a universal scaling relationship with shear modulus as
\begin{equation}
T_{\rm NI,0} - T_{\rm NI}({\mu^*}) \sim {\mu^*}^\frac{4}{5}
\end{equation}
Using the Landau-de Gennes form of nematic energy, Warner and Gelling predicted a different linear scaling as $T_{\rm NI,0} - T_{\rm NI}({\mu^*}) \sim {\mu^*}$.\cite{Warner1988} We think the $1/5$ difference in the scaling exponent comes from the fact that the Landau-de Gennes treatment only includes the leading-order terms of the nematic order parameter. Furthermore, the scaling relationship predicted by our theory facilitates access to LCEs with a targeted transition temperature. The complete information of transition temperature can be obtained by only measuring that of a few samples.

\subsection{Multiple Deformation Modes by Adjusting LC Grafting Density}

\begin{figure}[h] 
\centering
\includegraphics[width=1\textwidth]{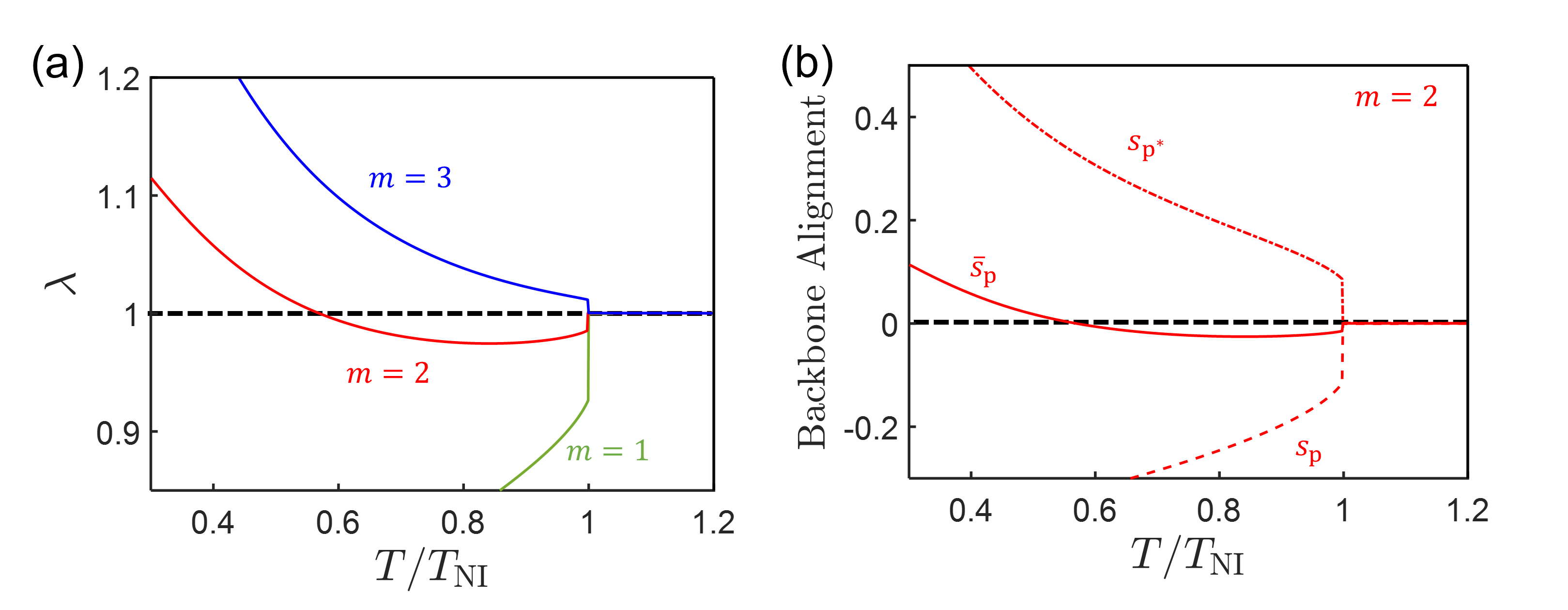}
\caption{Effect of LC grafting density on the spontaneous shape change for end-on SCLCE ($u_C = -0.25u_{\rm qq}$). LC grafting density is represented by $m$, the interval between two adjacent mesogens in the same chain. (a) Deformation $\lambda$ as a function of temperature for different $m$. Temperature is rescaled relative to the transition temperature $T_{\rm NI}$ for each $m$. (b) Order parameters of backbone alignment for LC-grafted ($s_{\rm p}$) and non-grafted ($s_{\rm p^*}$) backbone segments, as well as the average value ($\bar{s}_{\rm p}$) for the case of $m=2$. $r=1$, $u_{\rm pp}=u_{\rm p^*p^*}=u_{\rm pp^*}=0.07u_{\rm qq}$, $u_{\rm pq}=u_{\rm p^*q}=0.2u_{\rm qq}$, and $\mu^* = 0.5$.} 
\label{figure:figure5}
\end{figure}

The competition between the global and local couplings in SCLCEs and its resulting shape change can be modulated by varying LC grafting density. In our freely jointed chain model, LC grafting density is represented by the parameter $m$ which controls the interval between two adjacent mesogens in the same chain. Figure \ref{figure:figure5}a shows the effect of LC grafting density on the spontaneous shape change of end-on SCLCE ($u_C = -0.25u_{\rm qq}$). Both the fully grafted case ($m=1$) and the sparsely grafted case ($m=3$) exhibit a single deformation mode: the former is always contraction due to the dominant local coupling effect, whereas the latter is always elongation controlled by the global coupling. For the case of intermediate LC grafting density ($m=2$), it is interesting to note that the LCE has dual deformation modes. The LCE contracts as temperature decreases just below $T_{\rm NI}$, which is followed by a subsequent re-elongation as temperature further decreases. Such non-monotonic shape change sheds light on potential applications of LCEs as smart actuators with multiple responses to external stimuli and shape memory materials with diverse shape patterns.\cite{Ko2017}

The multiple deformation modes modulated by LC grafting density originate from the fact that the coupling effects experienced by the LC-grafted (${\rm p}$) and non-grafted (${\rm p^*}$) backbone segments are different. While the global coupling associated with the nematic field affects both types of backbone segments, the local coupling due to the chemical hinge only affects the directly grafted ones. As shown in Fig \ref{figure:figure5}b, non-grafted segments are aligned parallel to the director ($s_{\rm p^*}>0$), whereas the LC-grafted ones prefer a perpendicular alignment ($s_{\rm p}<0$) due to the tight hinge. Compared to $s_{\rm p}$, $s_{\rm p^*}$ is more sensitive to the decrease of temperature, particularly showing a more rapid change at low temperatures. This is because the parallel alignment of $\rm p^*$ concentrates only in one direction; however, the alignment of $\rm p$ is uniformly distributed in the entire plane perpendicular to the director. The overall effect is a non-monotonic change in the average backbone alignment $\bar{s}_{\rm p}$ from negative to positive, leading to the dual deformation modes as indicated by Eq. \ref{eq:deformation_scaler}.

\begin{figure}[h] 
\centering
\includegraphics[width=0.7\textwidth]{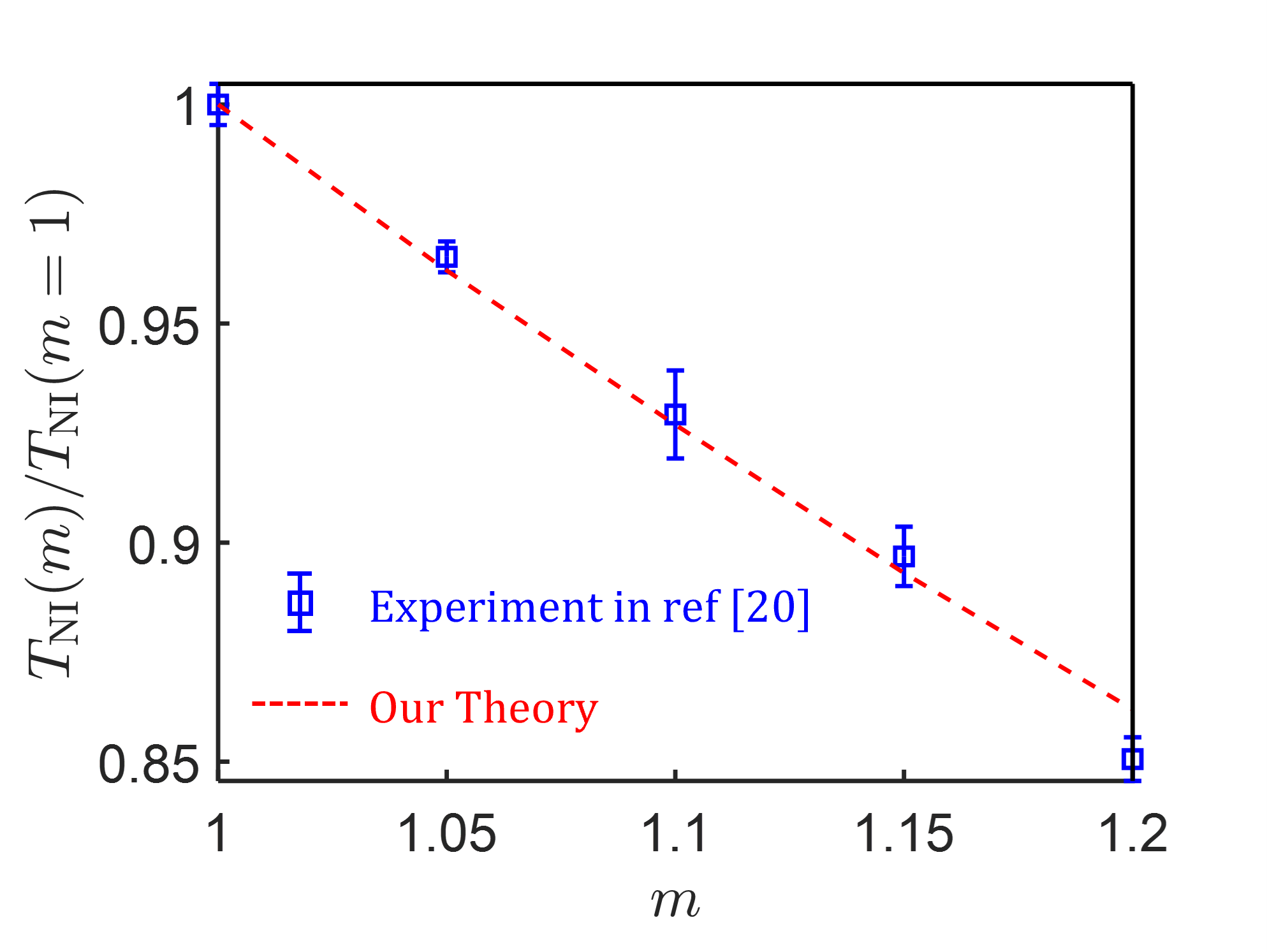}
\caption{Effect of the LC fraction on the nematic-isotropic transition temperature for MCLCE. The LC fraction is represented by $m$, where $m=1$ denotes the LC fully loaded case without any non-mesogenic segment. Our theoretical predictions using $\mu^*=0.3$ are compared to experimental data adopted from ref. [20].} 
\label{figure:figure6}
\end{figure}

Recent experiments showed a significant impact of liquid crystal loading on the phase transition temperature and actuation strain.\cite{Barnes2022, Saed2017soft, Saed2019} By replacing part of mesogens on the backbone of MCLCE with non-mesogenic molecules, Barens \textit{et al.} found a linear reduction of the transition temperature with respect to the LC fraction.\cite{Barnes2022} To validate our theory, we compare the theoretical predictions with the experimental results of Barens \textit{et al.}\cite{Barnes2022} As shown in Fig. \ref{figure:figure6}, our theory quantitatively captures the experimental data using $\mu^*=0.3$. The reduction of $T_{\rm NI}$ as $m$ increases is due to the dilution of mesogens by non-mesogenic molecules.

\subsection{Deformation and Critical Behavior under Applied Stress}

\begin{figure}[h] 
\centering
\includegraphics[width=1\textwidth]{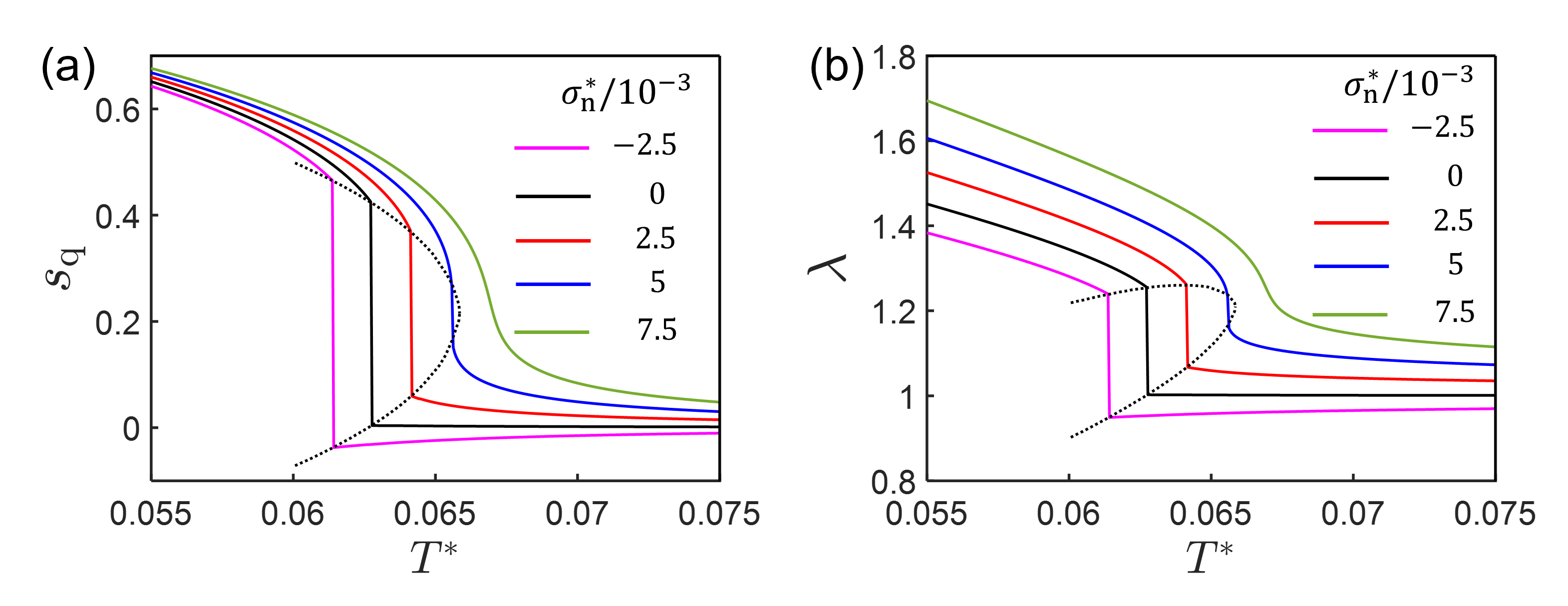}
\caption{Effect of applied nominal stress along the director on (a) mesogen nematic ordering $s_{\rm q}$ and (b) network deformation $\lambda$ for side-on SCLCEs ($u_C = 0.25 u_{\rm qq}$). $\sigma_{\rm n}^* = \sigma_{\rm n}/(2/3 U_{\rm qq})$ is the dimensionless nominal stress. The dotted lines denote the binodal envelope. $m=1$, $r=1$, $u_{\rm pp}=0.07u_{\rm qq}$, $u_{\rm pq}=0.2u_{\rm qq}$, and $\mu^*=0.5$.} 
\label{figure:figure7}
\end{figure}

LCE-based materials such as artificial muscles, often work under the condition of applied external stress. Such applied stress deforms the elastic network, changes the alignment of backbone segments, and hence alters the nematic ordering of SCLCEs via the local and global coupling effects. The nematic ordering in turn affects the deformation, leading to the deviation from classical rubber elasticity. Our self-consistent field theory enables a systematic inclusion of these effects in a unified framework. Figure \ref{figure:figure7} shows the nematic order parameter of mesogens $s_{\rm q}$ and deformation $\lambda$ for side-on SCLCEs ($u_C = 0.25 u_{\rm qq}$) under different nominal stress $\sigma_{\rm n}$. As shown in Fig. \ref{figure:figure7}a, nematic ordering is enhanced by the elongational stress ($\sigma_{\rm n}>0$) along the director, manifested by the increase of both transition temperature and order parameter. In contrast, the nematic ordering is suppressed by the compressive stress ($\sigma_{\rm n}<0$). Compared to the force-free counterpart, the mesogens in SCLCE show pre-ordering at temperatures higher than the transition point, which is known as the paranematic phase.\cite{Lebar2005,Cordoyiannis2009} Furthermore, the two order parameters belonging to the coexistent phases at the transition point approach each other as $\sigma_{\rm n}$ increases.  At the critical stress, they eventually merge into a single point. Beyond that point, the transition from the paranematic to nematic phase occurs in a continuous manner, i.e., supercritical behavior. The trend of $\lambda$ shown in Fig. \ref{figure:figure7}b follows a similar behavior as $s_{\rm q}$.

It has been observed in experiments that the first-order feature of the nematic phase transition in uncrosslinked LCP can degrade into a continuous one upon crosslinking.\cite{warner2007, Disch1994, Cordoyiannis2007, Cordoyiannis2009} Combining the results in Fig. \ref{figure:figure2} and Fig. \ref{figure:figure7}, we demonstrate that the crosslinking alone will not change the first-order nature of the phase transition. Instead, we suggest that this degradation can be explained by the residual stress retained in the elastomer when stretching a polydomain LCE to prepare a monodomain LCE at the second stage of the widely adopted two-step crosslinking procedure. \cite{Disch1994, warner2007, Herbert2022}

\begin{figure}[h] 
\centering
\includegraphics[width=0.7\textwidth]{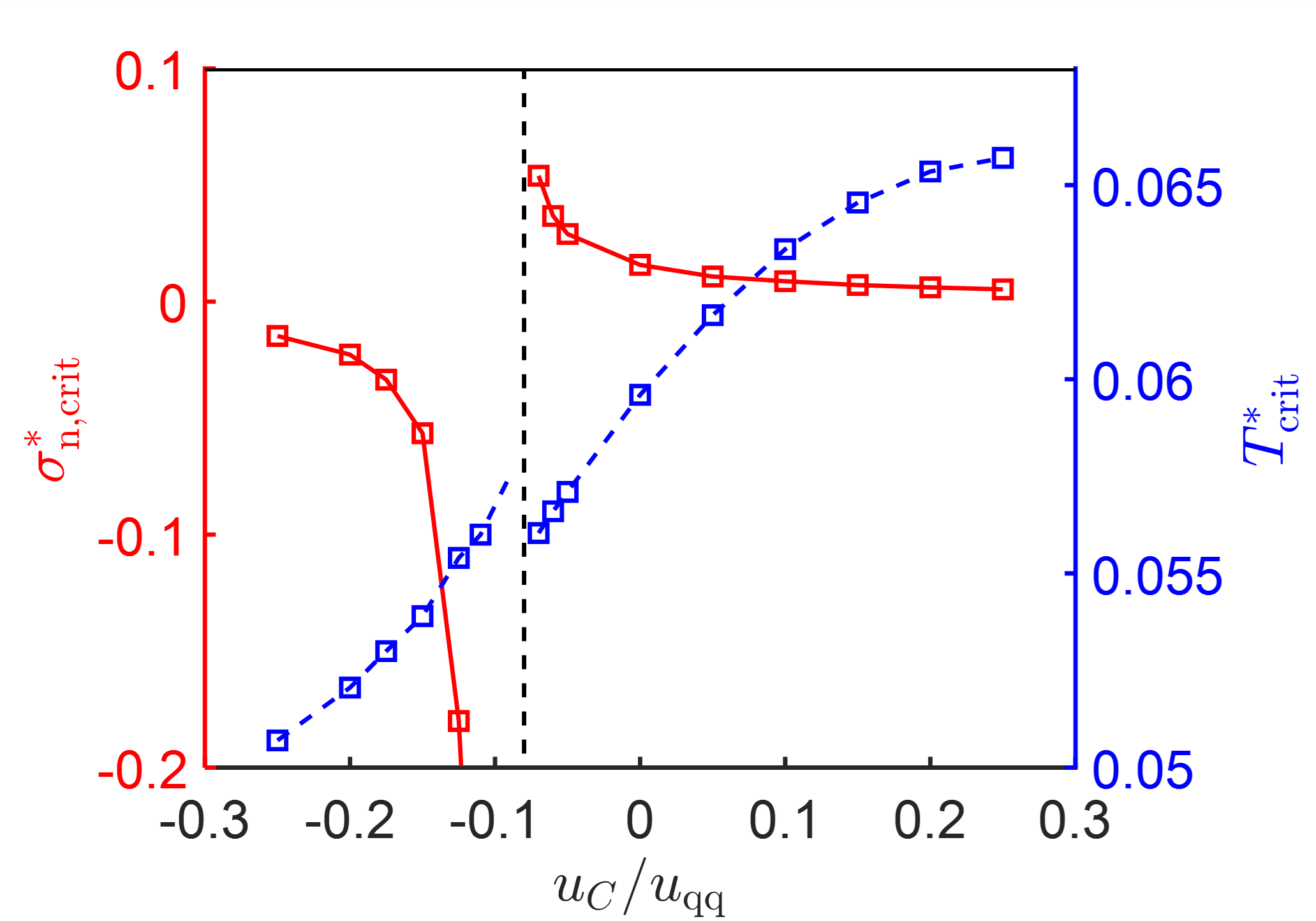}
\caption{Effect of local chemical hinge on the critical behavior of SCLCE. The critical nominal stress $\sigma_{\rm n,crit}^*$ and critical transition temperature $T_{\rm crit}^*$ are plotted as a function of the local coupling parameter $u_C$. The vertical dashed line denotes the discontinuity of $T_{\rm crit}^*$ and the singularity of $\sigma_{\rm n,crit}^*$ at the particular $u_C = -0.08u_{\rm qq}$. $m=1$, $r=1$, $u_{\rm pp}=0.07u_{\rm qq}$, $u_{\rm pq}=0.2u_{\rm qq}$, and $\mu^*=0.5$.} 
\label{figure:figure8}
\end{figure}

To predictably control the ordering and deformation of LCEs at various working conditions, it is important to understand the relationship between the critical point and the network structures. Figure \ref{figure:figure8} summarizes the effect of local chemical hinge on the critical applied nominal stress $\sigma_{\rm n,crit}$ and the corresponding critical transition temperature $T_{\rm crit}$. $T_{\rm crit}$ increases with $u_C$ but shows a discontinuity at the critical hinge value, i.e. $u_C=-0.08u_{\rm qq}$. On the other hand, $\sigma_{\rm n, crit}$ changes non-monotonically, and even exhibits a singularity at the critical $u_C$. For tightly hinged side-on SCLCE (a very positive $u_C$), only a small elongational stress is sufficient to bring the system to the supercritical regime. Similar small compressive stress is needed for the case of tightly hinged end-on SCLCE (a very negative $u_C$). As the hinge is loosened (i.e., the magnitude of $u_C$ decreases), the magnitude of the stress diverges towards the opposite limit near the singularity. At $u_C = -0.08u_{\rm qq}$, the global coupling effect is exactly canceled by the local coupling, such that mesogens are effectively decoupled from backbone segments. Therefore, no matter how large the network is deformed by the strong applied stress, it does not have any effect on the nematic ordering of mesogens, and the first-order transition will never degrade into a continuous one in this critical case.

\subsection{Mechanical Response to Applied Stress}

Understanding the stress-strain relationship plays is crucial for the application of rubbery materials.
In LCEs, the nematic ordering causes deviation in the mechanical response from the classical rubber elasticity. The problem becomes more complicated due to the interplay between mechanical deformation and nematic phase transition: the response is different between the nematic and isotropic phases. 

\begin{figure}[h] 
\centering
\includegraphics[width=1\textwidth]{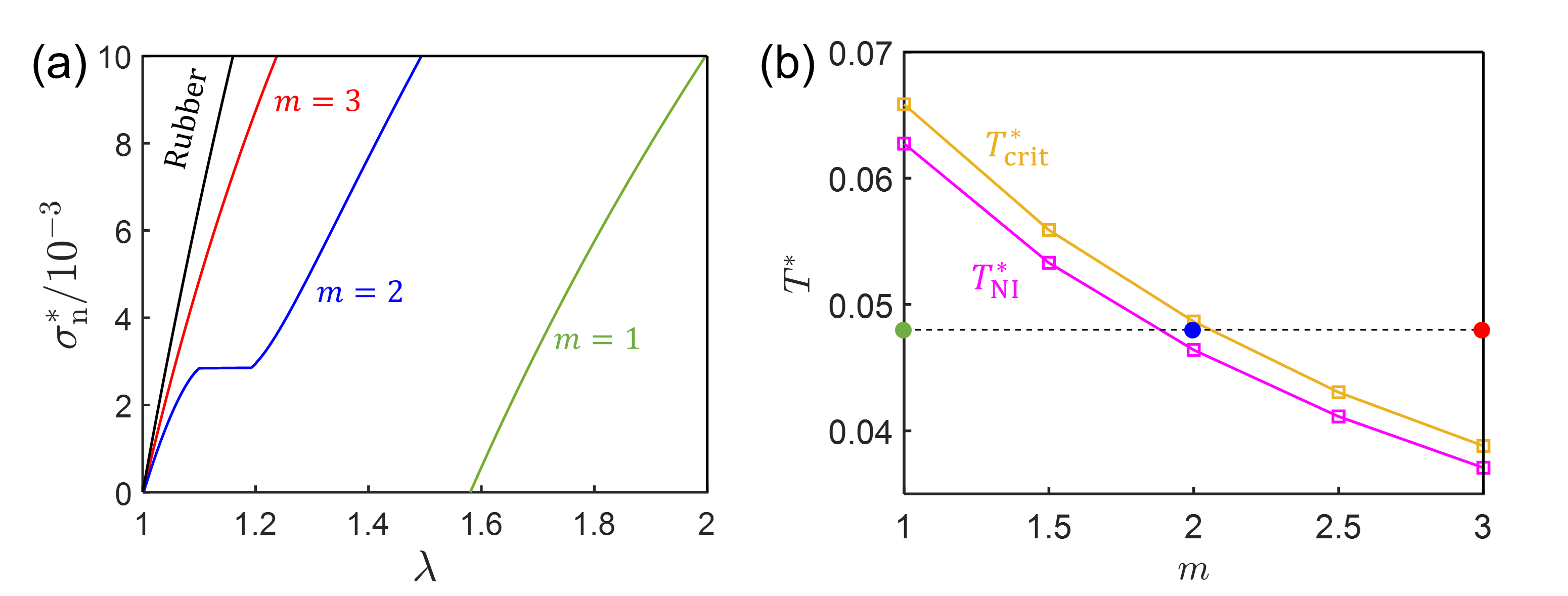}
\caption{Mechanical response of SCLCEs ($u_C = 0.25 u_{\rm qq}$) with different LC grafting density at the same fixed temperature. LC grafting density is represented by $m$, the interval between two adjacent mesogens in the same chain. (a) Rescaled nominal stress $\sigma_{\rm n}^*$ as a function of network deformation $\lambda$ for different $m$ at the working temperature $T^*=0.048$. (b) Rescaled nematic-isotropic transition temperature $T_{\rm NI}^*$ and critical temperature $T_{\rm c}^*$ as a function of $m$. The horizontal dashed line denotes the working temperature $T^*=0.048$. The red, blue, and green points locate the case of $m=1$, $m=2$, and $m=3$, respectively. $r=1$, $u_{\rm pp}=u_{\rm p^*p^*}=u_{\rm pp^*}=0.07u_{\rm qq}$, $u_{\rm pq}=u_{\rm p^*q}=0.2u_{\rm qq}$, and $\mu^* = 0.5$.} 
\label{figure:figure9}
\end{figure}

Figure \ref{figure:figure9}a plots the nominal stress as a function of deformation for side-on SCLCEs ($u_C = 0.25u_{\rm qq}$) with various LC grafting densities at the same fixed temperature. The behavior of isotropic rubber $\sigma_{\rm n} = \mu(\lambda - 1/\lambda^2)$ is also presented as a reference. The mechanical responses can be classified into three different types depending on the relation of the working temperature in comparison with $T_{\rm NI}$ and $T_{\rm crit}$ of the LCE. For the first case as the sparsely grafted LCE ($m=3$), the working temperature chosen is higher than $T_{\rm crit}$ (red point in Fig. \ref{figure:figure9}b), the LCE is always in the supercritical state during the deformation. The $\sigma_{\rm n}$-$\lambda$ curve is similar to that of the classical rubber but with a lower modulus. The LCE is easier to be deformed even in its paranematic state, because deformation is promoted by the alignment of backbone segments. For the second case as the LCE with intermediate grafting density ($m=2$), the working temperature is located between $T_{\rm NI}$ and $T_{\rm crit}$ (blue point in Fig. \ref{figure:figure9}b). In the regime of small deformation, the LCE remains in its paranematic state and is softer than the case of $m=3$. The stronger nematic ordering due to the increase of LC loading enhances backbone alignment as a result of the coupling effects. As $\sigma_{\rm n}$ becomes large enough, the stress triggers the discontinuous transition to the nematic phase, leading to a plateau in the $\sigma_{\rm n}$-$\lambda$ curve. After the transition, the deformation becomes easier as the elastomer is stretched in the nematic state. For the third case as the fully grafted LCE ($m=1$), the working temperature is lower than $T_{\rm NI}$ (green point in Fig. \ref{figure:figure9}b). The elastomer has a spontaneous deformation prior to the applied stress. The initial plateau at $\sigma_{\rm n}=0$ is followed by the stretch of the nematic elastomer.

Our results elucidate that the pattern of the mechanical response can be well controlled by the rational design of the LCE structure. In the current work, we predict that the plateau behavior during stretching originates from the discontinuous deformation accompanied by the first-order nematic phase transition. We note that similar plateau signature in the stress-strain curve of LCE-based materials can also be found in other cases. One example is the soft elasticity with respect to the director rotation if the stress is applied perpendicularly to the original nematic director.\cite{Warner1994} Another example is the re-alignment of local directors when stretching a polydomain LCE into a monodomain one.\cite{Schatzle1989,Barnes2022}

\section{Conclusions}

In this work, we develop a self-consistent field theory which provides a unified description of both main-chain and side-chain LCEs at the molecular level. Molecular features, such as density of elastic strands, the strength and architecture of chemical hinge, and LC grafting density are systematically included. Our theory fully captures the coupling between nematic ordering, backbone alignment and network deformation. The phase behavior and shape deformation both without and with the applied external stress are described in a single framework. We examine the effects of molecular structures on the nematic ordering and spontaneous shape change under the force-free condition. Compared to the uncrosslinked LCP, crosslinking is found to suppress the nematic ordering by reducing both the nematic order parameter and transition temperature as a result of the elastic energy stored during network deformation. Our results suggest that less crosslinked LCE achieves higher work capacity. While side-on SCLCEs always prefer elongational deformation, our theory captures the experimental observation that the shape change of end-on SCLCEs can be either elongation or contraction, determined by the competition between the global coupling associated with the nematic field and the local coupling due to the chemical hinge.
We elucidate a scaling relationship between the nematic-isotropic transition temperature and the shear modulus as $T_{\rm NI,0} - T_{\rm NI}({\mu^*}) \sim {\mu^*}^\frac{4}{5}$, which is universal for both MCLCE and SCLCEs with different hinge architectures. Furthermore, we show that the pattern of shape change can be well controlled by adjusting the LC grafting density. For end-on SCLCE with intermediate LC grafting density, the network deformation exhibits dual modes: contraction at higher temperatures and elongation at lower temperatures. The dependence of the transition temperature on LC grafting density predicted by our theory is in quantitive agreement with experimental data reported in previous literature. 

The phase behavior and mechanical response of LCEs under applied stress depend on the direction of the force and the nematic director. Nematic ordering is enhanced for side-on SCLCEs under elongational stress. The first-order transition eventually degrades to a continuous manner beyond the critical stress. We suggest that the continuous transition for LCEs observed in experiments can be explained by the residual stress retained in the elastomer when they are prepared via a two-step crosslinking procedure. At a critical strength of the local hinge, the local and global couplings cancel each other, leading to a discontinuity in the critical temperature and a singularity in the critical stress. Furthermore, the mechanical response of the LCEs deviates significantly from the classical rubber elasticity. The stress-deformation curve shows different patterns depending on the working temperature relative to the transition temperature and critical temperature. A plateau appears accompanied by the first-order nematic phase transition. 

Although the current work focuses on a homogeneous network with uniaxial deformation along the director, our theory is universal for LCEs with spatial inhomogeneity and under any form of stress. Our theory can be easily generalized to more complex systems, for example, heteropolymer networks consisting of multi-types of mesogens and LCEs swollen in isotropic or nematic solvents. Our theory can also incorporate the effect of other types of external stimuli, such as light, electric and magnetic fields. Modeling the phase and mechanical behaviors of LCEs at the molecular level is crucial for the establishment of their structure-property relationship. Our work is thus an important step towards the predictable design of LCE-based materials.

\begin{acknowledgement}

R. W. acknowledges the support from the University of California, Berkeley. This research used the computational resources provided by the Kenneth S. Pitzer Center for Theoretical Chemistry.

\end{acknowledgement}

\begin{suppinfo}

Detail derivation of self-consistent field theory for liquid crystal elastomers; numerical details for solving the self-consistent equations. This material is available free of charge via the
Internet at http://pubs.acs.org/.

\end{suppinfo}


\bibliography{refs}

\end{document}

%\linenumbers
\title[]
{Supporting Information: \\
A Molecular Theory for Liquid Crystal Elastomers: Nematic Ordering, Shape Deformation and Mechanical Response}

\author{Luofu Liu}
\affiliation{Department of Chemical and Biomolecular Engineering, University of California Berkeley, Berkeley, California 94720, United States}

\author{Rui Wang}
\email {ruiwang325@berkeley.edu}\affiliation{Department of Chemical and Biomolecular Engineering, University of California Berkeley, Berkeley, California 94720, United States}
\affiliation{Materials Sciences Division, Lawrence Berkeley National Lab, Berkeley, California 94720, United States}

\renewcommand*{\citenumfont}[1]{S#1}
\renewcommand*{\bibnumfmt}[1]{(S#1)}

\renewcommand{\theequation}{S\arabic{equation}}
\renewcommand{\thefigure}{S\arabic{figure}}

\maketitle

\newpage

\section{I. Derivation of the Self-consistent Field Theory}

In this section, we give a detailed derivation of the key equations in the SCFT for liquid crystal elastomers (LCEs). As discussed in the main text, we start by defining the microscopic operators. The density operators are given by
\begin{equation}\label{eq:density_operators}
\begin{aligned}
    &\hat{\phi}_{\rm p}(\bm{r}) = v_{\rm{p}}\sum_{\alpha=1}^{n}\sum_{j=1}^N\delta(\bm{r} - \bm{r}_{\alpha,(j-1)m+1})\\
    &\hat{\phi}_{\rm p^*}(\bm{r}) = v_{\rm{p}}\sum_{\alpha=1}^{n}\sum_{j=1}^N\sum_{k=2}^m\delta(\bm{r} - \bm{r}_{\alpha,(j-1)m+k})\\
    &\hat{\phi}_{\rm q}(\bm{r}) = v_{\rm{q}}\sum_{\alpha=1}^{n}\sum_{j=1}^N\delta(\bm{r} - \bm{r}_{\alpha,(j-1)m+1})
\end{aligned}
\end{equation}
where $\rm p$, $\rm p^*$, $\rm q$ refer to the LC-grafted backbone segment, non-grafted segment, and mesogen, respectively. $v_{\rm p}$ and $v_{\rm q}$ are the occupied volume of backbone segment and mesogen. The position vector of the $k$-th segment in the $j$-th repeating section on the $\alpha$-th chain is given by $\bm{r}_{\alpha,(j-1)m+k}=\bm{R}_{\alpha,0} + b\sum_{k=1}^{(j-1)m+k-1}\bm{p}_{\alpha,k}+(b/2)\bm{p}_{\alpha,(j-1)m+k}$ with $\bm{R}_{\alpha,0}$ the position of one of the chain-end. The nematic operators are given by
\begin{equation}\label{eq:orderparam_operators}
\begin{aligned}
    &\hat{\bm{S}}_{\rm p}(\bm{r}) = v_{\rm{p}}\sum_{\alpha=1}^{n}\sum_{j=1}^N\delta(\bm{r} - \bm{r}_{\alpha,(j-1)m+1})\left[\bm{p}_{\alpha,(j-1)m+1}\otimes\bm{p}_{\alpha,(j-1)m+1}-\frac{1}{3}\bm{I}\right]\\
    &\hat{\bm{S}}_{\rm p^*}(\bm{r}) = v_{\rm{p}}\sum_{\alpha=1}^{n}\sum_{j=1}^N\sum_{k=2}^m\delta(\bm{r} - \bm{r}_{\alpha,(j-1)m+k})\left[\bm{p}_{\alpha,(j-1)m+k}\otimes\bm{p}_{\alpha,(j-1)m+k}-\frac{1}{3}\bm{I}\right]\\
    &\hat{\bm{S}}_{\rm q}(\bm{r}) = v_{\rm{q}}\sum_{\alpha=1}^{n}\sum_{j=1}^N\delta(\bm{r} - \bm{r}_{\alpha,(j-1)m+1})\left[\bm{q}_{\alpha,(j-1)m+1}\otimes\bm{q}_{\alpha,(j-1)m+1}-\frac{1}{3}\bm{I}\right]
\end{aligned}
\end{equation}
where $\bm{I}$ stands for the identity tensor. 

The Hamiltonian of the system, including global nematic interactions, local chemical hinge, and network elastic energy, is given by
\begin{equation} \label{eq:total_H}
\begin{aligned}
\mathcal{H}_{\rm} =& -\frac{U_{\rm pp}}{2}\int {\rm d}\bm{r}\hat{\bm{S}}_{\rm p}(\bm{r}):\hat{\bm{S}}_{\rm p}(\bm{r}) -\frac{U_{\rm p^{*}p^{*}}}{2}\int {\rm d}\bm{r}\hat{\bm{S}}_{\rm p^*}(\bm{r}):\hat{\bm{S}}_{\rm p^*}(\bm{r})  -\frac{U_{\rm qq}}{2}\int {\rm d}\bm{r}\hat{\bm{S}}_{\rm q}(\bm{r}):\hat{\bm{S}}_{\rm q}(\bm{r})\\
&-U_{\rm pp^*}\int {\rm d}\bm{r}\hat{\bm{S}}_{\rm p}(\bm{r}):\hat{\bm{S}}_{\rm p^*}(\bm{r}) -U_{\rm pq}\int {\rm d}\bm{r}\hat{\bm{S}}_{\rm p}(\bm{r}):\hat{\bm{S}}_{\rm q}(\bm{r}) -U_{\rm p^*q}\int {\rm d}\bm{r}\hat{\bm{S}}_{\rm p^*}(\bm{r}):\hat{\bm{S}}_{\rm q}(\bm{r})\\
&- \sum_{\alpha=1}^{n}\sum_{j=1}^N u_C(\bm{p}_{\alpha, (j-1)m+1}\cdot \bm{q}_{\alpha, (j-1)m+1})^2 + \frac{1}{2}\mu \int {\rm d}\bm{r} \rm {Tr} [\bm{l}_0 \bm{\lambda}^{\rm T}(\bm{r}) \hat{\bm{l}}^{-1}(\bm{r}) \bm{\lambda}(\bm{r}) ]
\end{aligned}
\end{equation}
where $U_{\kappa \kappa'}$ is the nematic interaction parameter between species $\kappa$ and $\kappa'$ ($\kappa,\kappa'={\rm p},{\rm p^*},{\rm q}$), $u_C$ is the local coupling strength, $\mu$ is the linear shear modulus of the network, $\bm{l}_0$ is the step length tensor at the as-prepared state, $\hat{\bm{l}}$ is the (particle-based) step length tensor for the deformed state, $\bm{\lambda}$ is the deformation gradient tensor. $\rm{Tr}$ refers to the trace of the tensor. $\hat{\bm{l}}$ can be related to the orientations of backbone segments as
\begin{equation}
\hat{\bm{l}}(\bm{r}) = b\left\{\bm{I} + \frac{1}{m} \left[\frac{3\hat{\bm{S}}_{\rm p}(\bm{r})}{\hat{\phi}_{\rm p}(\bm{r})} \right] + \frac{m-1}{m} \left[\frac{3\hat{\bm{S}}_{\rm p^*}(\bm{r})}{\hat{\phi}_{\rm p^*}(\bm{r})} \right] \right\}
\end{equation}

For LCEs under an applied nominal stress $\bm{\sigma}_{\rm n}$, the partition function of the system is then calculated by integrating all configurations of segments and mesogens as well as different deformations:
\begin{equation} \label{eq:total_partition_function}
\begin{aligned}
Z &= \int {\rm D}\bm{\lambda}\exp \left[
\beta \int {\rm d}\bm{r} \bm{\sigma}_{\rm n}:\bm{\lambda}(\bm{r}) \right] 
\prod_{\alpha=1}^{n}\left(
\int {\rm d}\bm{R}_{\alpha,0} \prod_{j=1}^N \prod_{k=1}^m \int {\rm d} \bm{p}_{\alpha, (j-1)m+k} \cdot \prod_{j=1}^N \int {\rm d} \bm{q}_{\alpha, (j-1)m+1} \right)\\
&\prod_{\bm{r}} \delta\left[\hat{\phi}_{\rm p}(\bm{r}) +\hat{\phi}_{\rm p^*}(\bm{r}) + \hat{\phi}_{\rm q}(\bm{r}) -1 \right] \cdot \prod_{\bm{r}}\delta \left[ \det \bm{\lambda}(\bm{r}) - 1 \right] \exp(-\beta \mathcal{H})
\end{aligned}
\end{equation}
The functional integral over $\bm{\lambda}$ results from the transform of ensemble from constant deformation to constant applied stress. The first $\delta$-functional enforces the local incompressibility, whereas the second one originates from the volume conservation of incompressible materials under deformation. We perform the standard self-consistent field approach. The field-based partition function is obtained by introducing auxiliary fields via identity transformations.
The following identities will be inserted into the partition function:
\begin{equation}
1 \equiv \int {\rm D} \bm{S}_{ \kappa} \prod_{r}\delta\left[\bm{S}_{\kappa}(\bm{r}) - \hat{\bm{S}}_{\kappa}(\bm{r}) \right] = \int {\rm D} \bm{S}_{\kappa} \int {\rm D} \bm{\Lambda}_{\kappa} \exp \left\{ i \int {\rm d} \bm{r} \bm{\Lambda}_{\kappa}(\bm{r}):\left[\bm{S}_{\kappa}(\bm{r}) - \hat{\bm{S}}_{\kappa}(\bm{r}) \right] \right\}, \ \ \kappa = {\rm p}, \ {\rm p^*}, \ \rm{q}
\end{equation}
and the $\delta$-functionals are treated by Fourier transform:
\begin{equation}
\begin{aligned}
&\prod_{\bm{r}} \delta\left[\hat{\phi}_{\rm p}(\bm{r}) +\hat{\phi}_{\rm p^*}(\bm{r}) + \hat{\phi}_{\rm q}(\bm{r}) -1 \right] = \int {\rm D} \xi \exp \left\{i\int {\rm d} \bm{r} \xi(\bm{r}) \left[\hat{\phi}_{\rm p}(\bm{r}) +\hat{\phi}_{\rm p^*}(\bm{r}) + \hat{\phi}_{\rm q}(\bm{r}) -1 \right] \right\} \\
&\prod_{\bm{r}}\delta \left[ \det \bm{\lambda}(\bm{r}) - 1 \right] =  \int {\rm D} \eta \exp \left\{ i \int {\rm d} \bm{r} \eta(\bm{r}) \left[ \det \bm{\lambda}(\bm{r}) - 1 \right]    \right\}
\end{aligned}
\end{equation}
then the field-based partition function is given by
\begin{equation}\label{eq:field_based_partition_function}
Z = \int {\rm D}\bm{S}_{\rm p} \int {\rm D}\bm{\Lambda}_{\rm p} \int {\rm D}\bm{S}_{\rm p^*} \int {\rm D}\bm{\Lambda}_{\rm p^*} \int {\rm D}\bm{S}_{\rm q} \int {\rm D}\bm{\Lambda}_{\rm q} \int {\rm D}\xi \int {\rm D}\eta \int {\rm D}\bm{\lambda} \exp(-\beta L)
\end{equation}
The ``action" $L$ is
\begin{equation}\label{eq:free_energy_complete}
\begin{aligned}
\beta L =& i \int {\rm d} \bm{r} \xi(\bm{r}) - i \int {\rm d} \bm{r} \eta (\bm{r}) \left[\det \bm{\lambda}(\bm{r}) - 1  \right] \\ 
&-\frac{\beta U_{\rm pp}}{2}\int {\rm d}\bm{r}{\bm{S}}_{\rm p}(\bm{r}):{\bm{S}}_{\rm p}(\bm{r}) -\frac{\beta U_{\rm p^{*}p^{*}}}{2}\int {\rm d}\bm{r}{\bm{S}}_{\rm p^*}(\bm{r}):{\bm{S}}_{\rm p^*}(\bm{r})  -\frac{\beta U_{\rm qq}}{2}\int {\rm d}\bm{r}{\bm{S}}_{\rm q}(\bm{r}):{\bm{S}}_{\rm q}(\bm{r})\\
&-\beta U_{\rm pp^*}\int {\rm d}\bm{r}{\bm{S}}_{\rm p}(\bm{r}):{\bm{S}}_{\rm p^*}(\bm{r}) -\beta U_{\rm pq}\int {\rm d}\bm{r}{\bm{S}}_{\rm p}(\bm{r}):{\bm{S}}_{\rm q}(\bm{r}) -\beta U_{\rm p^*q}\int {\rm d}\bm{r}{\bm{S}}_{\rm p^*}(\bm{r}):{\bm{S}}_{\rm q}(\bm{r})\\
&-i \int {\rm d} \bm{r} \left[\bm{\Lambda}_{\rm p}(\bm{r}):\bm{S}_{\rm p}(\bm{r}) + \bm{\Lambda}_{\rm p^*}(\bm{r}):\bm{S}_{\rm p^*}(\bm{r}) + \bm{\Lambda}_{\rm q}(\bm{r}):\bm{S}_{\rm q}(\bm{r})\right] + \frac{1}{2} \beta \mu \int {\rm d} \bm{r} {\rm Tr} \left[ \bm{l}_0 \bm{\lambda}^{\rm T}(\bm{r}) {\bm l}^{-1}(\bm{r}) \bm{\lambda}(\bm{r}) \right] \\
& - n \ln Q\left[ \bm{\Lambda}_{\rm p}(\bm{r}), \bm{\Lambda}_{\rm p^*}(\bm{r}), \bm{\Lambda}_{\rm q}(\bm{r}), \xi(\bm{r})   \right] -\beta \int {\rm d} \bm{r} \bm{\sigma}_{\rm n}(\bm{r}): \bm{\lambda}(\bm{r})
\end{aligned}
\end{equation}
where $Q$ is the single-chain partition function, given by
\begin{equation}
\begin{aligned}
Q =& \prod_{j=1}^N \prod_{k=1}^m \int {\rm d} \bm{p}_{(j-1)m+k} \cdot \prod_{j=1}^N \int {\rm d} \bm{q}_{(i-1)m+1} \exp\left[\sum_{j=1}^N \beta u_C(\bm{p}_{ (j-1)m+1} \cdot \bm{q}_{ (j-1)m+1})^2 \right] \\
&\exp \left\{ \sum_{j=1}^N \left[iv_{\rm p}\xi(\bm{r}_{(j-1)m+1}) -iv_{\rm p} \bm{\Lambda}_{\rm p}(\bm{r}_{(j-1)m+1}) : \left(\bm{p}_{(j-1)m+1} \otimes \bm{p}_{(j-1)m+1} -\frac{1}{3} \bm{I} \right)  \right] \right\} \\
&\exp \left\{ \sum_{j=1}^N \sum_{k=2}^m \left[iv_{\rm p}\xi(\bm{r}_{(j-1)m+k}) -iv_{\rm p} \bm{\Lambda}_{\rm p^*}(\bm{r}_{(j-1)m+k}) : \left(\bm{p}_{(j-1)m+k} \otimes \bm{p}_{(j-1)m+k} -\frac{1}{3} \bm{I} \right)  \right] \right\} \\
&\exp \left\{ \sum_{j=1}^N \left[iv_{\rm q}\xi(\bm{r}_{(j-1)m+1}) -iv_{\rm q} \bm{\Lambda}_{\rm q}(\bm{r}_{(j-1)m+1}) : \left(\bm{q}_{(j-1)m+1} \otimes \bm{q}_{(j-1)m+1} -\frac{1}{3} \bm{I} \right)  \right] \right\}
\end{aligned}
\end{equation}

The direct calculation of the functional integration in Eq. \ref{eq:field_based_partition_function} is intractable due to the complicated coupling between fields. Here we use the saddle-point approximation to find the state that contributes most to the integration, i.e., the state with minimum free energy. This is achieved by taking the functional derivative of $F$ with respect to all field variables to be zero (or zero tensor). This step derives the following set of self-consistent equations describing the coupling between nematic ordering and shape deformation:
\begin{subequations}\label{eq:self_consistent_equations_SI}
\begin{equation}
 \bm{h}_{\rm p} (\bm{r}) \equiv i\bm{\Lambda}_{\rm p}(\bm{r}) = -\beta \left[ U_{\rm pp} \bm{S}_{\rm p}(\bm{r}) + U_{\rm pp^*}\bm{S}_{\rm p^*}(\bm{r}) + U_{\rm pq}\bm{S}_{\rm q}(\bm{r})+\frac{3\mu}{2 m \phi_{\rm p}(\bm{r})} b \bm{l}^{-1}(\bm{r})\bm{\lambda}(\bm{r})\bm{l}_0\bm{\lambda}^{\rm T}(\bm{r})\bm{l}^{-1}(\bm{r}) \right] \label{scfSI_a}
\end{equation}
\begin{equation}
\bm{h}_{\rm p^*} (\bm{r}) \equiv i\bm{\Lambda}_{\rm p^*}(\bm{r}) = -\beta \left[ U_{\rm pp^*} \bm{S}_{\rm p}(\bm{r}) + U_{\rm p^*p^*}\bm{S}_{\rm p^*}(\bm{r}) +  U_{\rm p^*q}\bm{S}_{\rm q}(\bm{r}) + \frac{3(m-1)\mu}{2 m \phi_{\rm p^*}(\bm{r})} b \bm{l}^{-1}(\bm{r})\bm{\lambda}(\bm{r})\bm{l}_0\bm{\lambda}^{\rm T}(\bm{r})\bm{l}^{-1}(\bm{r}) \right] \label{scfSI_b}
\end{equation}
\begin{equation}
\bm{h}_{\rm q} (\bm{r}) \equiv i\bm{\Lambda}_{\rm q}(\bm{r}) = -\beta \left[ U_{\rm pq} \bm{S}_{\rm p}(\bm{r}) + U_{\rm p^*q}\bm{S}_{\rm p^*}(\bm{r}) + U_{\rm qq}\bm{S}_{\rm q}(\bm{r}) \right] \label{scfSI_c} 
\end{equation}
\begin{equation}
 \bm{S}_{\rm p}(\bm{r}) = -\frac{n}{Q}\frac{\delta Q}{\delta \bm{h}_{\rm p}(\bm{r})}\label{scfSI_d} 
\end{equation}
\begin{equation}
 \bm{S}_{\rm p^*}(\bm{r}) = -\frac{n}{Q}\frac{\delta Q}{\delta \bm{h}_{\rm p^*}(\bm{r})} \label{scfSI_e}
\end{equation}
\begin{equation}
 \bm{S}_{\rm q}(\bm{r}) = -\frac{n}{Q}\frac{\delta Q}{\delta \bm{h}_{\rm q}(\bm{r})} \label{scfSI_f} 
\end{equation}
\begin{equation}
\bm{l}_0 \bm{\lambda}^{\rm T}(\bm{r})\bm{l}^{-1}(\bm{r}) + \bm{l}_0^{\rm T} \bm{\lambda}^{\rm T}(\bm{r})\bm{l}^{-\rm T}(\bm{r}) - \gamma(\bm{r})  \bm{\lambda}^{-1}(\bm{r}) -2\frac{\bm{\sigma}_{\rm n}(\bm{r})}{\mu}= \bm{0} \label{scfSI_g}
\end{equation}
\begin{equation}
\det \bm{\lambda}(\bm{r}) - 1 = 0 \label{scfSI_h}
\end{equation}
\begin{equation}
\phi_{\rm p}(\bm{r}) + \phi_{\rm p^*}(\bm{r}) + \phi_{\rm q}(\bm{r}) - 1 = 0 \label{scfSI_i}
\end{equation}
\end{subequations}
where $\gamma = \frac{2i \eta \det \bm{\lambda}}{\beta \mu}$.

The complete set of Eqs. \ref{eq:self_consistent_equations_SI} is applicable to problems where the densities and order parameters are spatially inhomogeneous. In the current work, we focus on the bulk phase behavior and shape change, so we assume that all these variables are not position-dependent. In this sense, the single-chain partition function can be simplified as 
\begin{equation}\label{eq:bulk_single_chain_partition_function}
\begin{aligned}
Q =& \prod_{j=1}^N \prod_{k=1}^m \int {\rm d} \bm{p}_{ (j-1)m+k} \cdot \prod_{j=1}^N \int {\rm d} \bm{q}_{ (i-1)m+1} \exp\left[\sum_{j=1}^N \beta u_C(\bm{p}_{ (j-1)m+1} \cdot \bm{q}_{ (j-1)m+1})^2 \right] \\
&\exp \left\{ \sum_{j=1}^N \left[\beta v_{\rm p} \left( U_{\rm pp} \bm{S}_{\rm p} + U_{\rm pp^*} \bm{S}_{\rm p^*} + U_{\rm pq} \bm{S}_{\rm q} + \frac{3\mu}{2m\phi_{\rm p}} b \bm{l}^{-1}\bm{\lambda}\bm{l}_0\bm{\lambda}^{\rm T}\bm{l}^{-1} \right) : \left(\bm{p}_{(j-1)m+1} \otimes \bm{p}_{(j-1)m+1} -\frac{1}{3} \bm{I} \right)  \right] \right\} \\
&\exp \left\{ \sum_{j=1}^N \sum_{k=2}^m \left[\beta v_{\rm p} \left( U_{\rm pp^*} \bm{S}_{\rm p} + U_{\rm p^*p^*} \bm{S}_{\rm p^*} + U_{\rm p^*q} \bm{S}_{\rm q} + \frac{3(m-1)\mu}{2m\phi_{\rm p^*}} b \bm{l}^{-1}\bm{\lambda}\bm{l}_0\bm{\lambda}^{\rm T}\bm{l}^{-1} \right) : \left(\bm{p}_{(j-1)m+k} \otimes \bm{p}_{(j-1)m+k} -\frac{1}{3} \bm{I} \right)  \right] \right\} \\
&\exp \left\{ \sum_{j=1}^N \left[\beta v_{\rm q} \left( U_{\rm pq} \bm{S}_{\rm p} + U_{\rm p^*q} \bm{S}_{\rm p^*} + U_{\rm qq} \bm{S}_{\rm q} \right) : \left(\bm{q}_{(j-1)m+1} \otimes \bm{q}_{(j-1)m+1} -\frac{1}{3} \bm{I} \right)  \right] \right\}
\end{aligned}
\end{equation}
and the tensorial order parameters are given by
\begin{subequations}
\begin{equation}
\bm{S}_{\rm p} = \frac{nv_{\rm p}}{V} \left \langle\sum_{j=1}^N\left(\bm{p}_{(j-1)m+1}\otimes\bm{p}_{(j-1)m+1} - \frac{1}{3} \bm{I} \right) \right \rangle
\end{equation}
\begin{equation}
\bm{S}_{\rm p^*} = \frac{nv_{\rm p}}{V} \left \langle\sum_{j=1}^N \sum_{k=2}^m\left(\bm{p}_{(j-1)m+k}\otimes\bm{p}_{(j-1)m+k} - \frac{1}{3} \bm{I} \right) \right \rangle
\end{equation}
\begin{equation}
\bm{S}_{\rm q} = \frac{nv_{\rm q}}{V} \left \langle\sum_{j=1}^N\left(\bm{q}_{(j-1)m+1}\otimes\bm{q}_{(j-1)m+1} - \frac{1}{3} \bm{I} \right) \right \rangle
\end{equation}
\end{subequations}
where $\langle ... \rangle$ stands for the ensemble average taken based on the Boltzmann factor evaluated from the exponent in Eq. \ref{eq:bulk_single_chain_partition_function}.

For simplicity, we only consider the nematic ordering and deformation along the $z$-axis. Then all the tensorial order parameters are diagonal and traceless. Under the standard definition, the scaler order parameters are given by
\begin{subequations}
\begin{equation}
s_{\rm p} = \frac{3}{2}\frac{\bm{S}_{{\rm p},zz}}{\phi_{\rm p}} =\frac{1}{N} \left \langle\sum_{j=1}^N P_2 \left(\cos \theta_{{\rm p},(j-1)m+1} \right) \right \rangle
\end{equation}
\begin{equation}
s_{\rm p^*} = \frac{3}{2}\frac{\bm{S}_{{\rm p^*},zz}}{\phi_{\rm p^*}} =\frac{1}{N(m-1)} \left \langle\sum_{j=1}^N \sum_{k=2}^m P_2 \left(\cos \theta_{{\rm p^*},(j-1)m+k} \right) \right \rangle
\end{equation}
\begin{equation}
s_{\rm q} = \frac{3}{2}\frac{\bm{S}_{{\rm q},zz}}{\phi_{\rm q}} =\frac{1}{N} \left \langle\sum_{j=1}^N P_2 \left(\cos \theta_{{\rm q},(j-1)m+1} \right) \right \rangle
\end{equation}
\end{subequations}
where $P_2(x) = \frac{3}{2}x^2 - \frac{1}{2}$ is the second Legendre polynomial, $\theta_\kappa$ is the angle between the nematic director and the the unit orientational vector $\kappa$. Under the uniaxial nematic ordering, both the deformation gradient $\bm{\lambda}$ and nomimal stress are diagonal: $\lambda_{zz} = \lambda$, $\lambda_{xx} = \lambda_{yy} = 1/\sqrt{\lambda}$; $\sigma_{{\rm n},zz}=\sigma_{\rm n}$, $\sigma_{{\rm n},xx}=\sigma_{{\rm n},yy}=0$. Eq. \ref{scfSI_g} can be further simplified to 
\begin{equation}\label{eq:spontaneous_deformation}
    \sigma_{\rm n} = \mu \left( \lambda \frac{1+2\bar{s}_{\rm p,0}}{1+2\bar{s}_{\rm p}} - \frac{1}{\lambda^2} \frac{1-\bar{s}_{\rm p,0}}{1-\bar{s}_{\rm p}}\right)
\end{equation}
where $\bar{s}_{\rm p} = \frac{1}{m} s_{\rm p} + \frac{m-1}{m} s_{\rm p^*}$ is the average backbone order parameter, and $\bar{s}_{\rm p,0}$ is $\bar{s}_{\rm p}$ at the as-prepared state. Specially, when $\bar{s}_{\rm p,0} = \bar{s}_{\rm p}=0$, we recover the stress-deformation relation for an isotropic rubber:
\begin{equation}
    \sigma_{\rm n} = \mu \left( \lambda - \frac{1}{\lambda^2} \right)
\end{equation}
Or, without applied stress (i.e., force-free state, $\sigma_{\rm n}=0$), we recover the well-known equation describing the spontaneous deformation in nematic LCEs:
\begin{equation}
\lambda = \left( \frac{1-{s}_{\rm p,0}}{1+2{s}_{\rm p,0}} \right)^{\frac{1}{3}} \left( \frac{1+2{s}_{\rm p}}{1-{s}_{\rm p}} \right)^\frac{1}{3}
\end{equation}

Further, the single-chain partition function can be further simplified and decoupled into the level of repeating sections (and further chain segments):
\begin{equation}
Q  = Q_{\rm S}^N, \ \ Q_{\rm S} = Q_{1}Q_2^{m-1}
\end{equation}
where $Q_1$ is the partition function of mesogen and the backbone segment attached to it, given by
\begin{equation}\label{eq:Q1}
\begin{aligned}
Q_1 = &\int {\rm d}\bm{p} \int {\rm d}\bm{q} \exp\left[ \beta u_C(\bm{p}\cdot\bm{q})^2 \right] \\
&\exp \left \{ \frac{m+r}{1+r} \left[  \beta u_{\rm pp}\phi_{\rm p}^2 s_{\rm p} + \beta u_{\rm pp^*}\phi_{\rm p} \phi_{\rm p^*}s_{\rm p^*}  + \beta u_{\rm pq}\phi_{\rm p} \phi_{\rm q}s_{\rm q} + \frac{\mu^*}{m} \left(\frac{\lambda^2(1+2\bar{s}_{\rm p,0})}{(1+2\bar{s}_{\rm p})^2} - \frac{1-\bar{s}_{\rm p,0}}{\lambda(1-\bar{s}_{\rm p})^2} \right)  \right] P_2(\cos \theta_{\rm p})\right\}\\
&\exp \left \{\frac{m+r}{1+r} \left[\beta u_{\rm pq}\phi_{\rm p}\phi_{\rm q} s_{\rm p} + \beta u_{\rm p^*q}\phi_{\rm p^*} \phi_{\rm q}s_{\rm p^*}  + \beta u_{\rm qq}\phi_{\rm q}^2 s_{\rm q} \right] P_2(\cos \theta_{\rm q})\right\}
\end{aligned}
\end{equation}
and $Q_2$ is the partition function of non-grafted backbone, given by
\begin{equation}\label{eq:Q2}
Q_2 = \int {\rm d} \bm{p}^* \exp \left\{ \frac{m+r}{1+r} \left[ \frac{\beta u_{\rm pp^*}\phi_{\rm p}\phi_{\rm p^*} s_{\rm p} + \beta u_{\rm p^*p^*}\phi_{\rm p^*}^2 s_{\rm p^*}  + \beta u_{\rm p^*q}\phi_{\rm p^*} \phi_{\rm q} s_{\rm q}}{m-1} + \frac{\mu^*}{m} \left( \frac{\lambda^2(1+2\bar{s}_{\rm p,0})}{(1+2\bar{s}_{\rm p})^2} - \frac{(1-\bar{s}_{\rm p,0})}{\lambda(1-\bar{s}_{\rm p})^2} \right) \right]P_2(\cos \theta_{\rm p^*})\right\}
\end{equation}
where $u_{\kappa \kappa'} = \frac{2}{3} U_{\kappa \kappa'}(v_{\rm p}+v_{\rm q})$ is the rescaled interaction strength between species $\kappa$ and $\kappa'$, $\mu^* = \beta \mu (v_{\rm p} + v_{\rm q})$ is the rescaled modulus, and $r=v_{\rm q}/v_{\rm p}$ is the volume ratio of mesogen and chain backbone. 

Following the assumption of system homogeneity and uniaxial deformation, the saddle-point free energy from Eq. \ref{eq:free_energy_complete} can be further simplified to the free energy per repeating section:
\begin{equation}\label{eq:section_free_energy}
\begin{aligned}
\beta F_{\rm S} = & \frac{m+r}{1+r} \left(\frac{1}{2}\beta u_{\rm pp} \phi_{\rm p}^2 s_{\rm p}^2 +  \frac{1}{2}\beta u_{\rm p^*p^*} \phi_{\rm p^*}^2 s_{\rm p^*}^2 + \frac{1}{2}\beta u_{\rm qq} \phi_{\rm q}^2 s_{\rm q}^2 + \beta u_{\rm p p^*}\phi_{\rm p}\phi_{\rm p^*}s_{\rm p}s_{\rm p^*} + \beta u_{\rm pq} \phi_{\rm p} \phi_{\rm q}s_{\rm p} s_{\rm q} + \beta u_{\rm p^*q}\phi_{\rm p^*}\phi_{\rm q}s_{\rm p^*} s_{\rm q}\right) \\
& - \ln Q_1 - (m-1)\ln Q_2  + \frac{m+r}{1+r}\frac{\mu^*}{2} \left[ \lambda^2(1+2\bar{s}_{\rm p,0}) \frac{1+4\bar{s}_{\rm p}}{(1+2\bar{s}_{\rm p})^2} + \frac{2}{\lambda}(1-\bar{s}_{\rm p,0}) \frac{1-2\bar{s}_{\rm p}}{(1-\bar{s}_{\rm p})^2} - \frac{2\sigma_{\rm n}}{\mu}\lambda \right]
\end{aligned}
\end{equation}
where $\sigma_{\rm n}^* = (3/2)\sigma_{\rm n}/U_{\rm qq}$ is the rescaled nominal stress. The nematic order parameters can be described based on the chain segment:
\begin{subequations}\label{eq:simplest_order_parameter}
\begin{equation}
s_{\rm p} = \left \langle P_2(\cos \theta_{\rm p}) \right \rangle = \frac{\int {\rm d}\bm{p} \int {\rm d}\bm{q} P_2(\cos \theta_{\rm p})\cdot I_{1} (\bm{p},\bm{q};\lambda, s_{\rm p}, s_{\rm p^*}, s_{\rm q})}{Q_1}
\end{equation}
\begin{equation}
s_{\rm q} = \left \langle P_2(\cos \theta_{\rm q}) \right \rangle = \frac{\int {\rm d}\bm{p} \int {\rm d}\bm{q} P_2(\cos \theta_{\rm q})\cdot I_{1} (\bm{p},\bm{q};\lambda, s_{\rm p}, s_{\rm p^*}, s_{\rm q})}{Q_1}
\end{equation}
\begin{equation}
s_{\rm p^*} = \left \langle P_2(\cos \theta_{\rm p^*}) \right \rangle = \frac{\int {\rm d}{\bm{p}^*} P_2(\cos \theta_{\rm p^*})\cdot I_{2} (\bm{p}^*;\lambda, s_{\rm p}, s_{\rm p^*}, s_{\rm q})}{Q_2}
\end{equation}    
\end{subequations}
where $I_1$ and $I_2$ refer to the integrand in the expressions of $Q_1$ and $Q_2$ in Eqs. \ref{eq:Q1} and \ref{eq:Q2}.

The Eqs. \ref{eq:spontaneous_deformation} and \ref{eq:simplest_order_parameter} constitute the simplest form of self-consistent equations for the homogeneous and uniaxial nematic SCLCEs. They can be numerically solved to obtain the coupled nematic ordering and spontaneous shape deformation behaviors. It is worth noting that the Eqs. \ref{eq:spontaneous_deformation} and \ref{eq:simplest_order_parameter} can equivalently be obtained by directly minimizing the free energy in Eq. \ref{eq:section_free_energy} with respect to $s_{\rm p}$, $s_{\rm p^*}$, $s_{\rm q}$ and $\lambda$.

\section{II. Numerical Details}

When evaluating the nematic order parameters $s_{\rm p}$ and $s_{\rm q}$, the calculation of the integrals in the ensemble average $\langle P_2(\cos\theta) \rangle$ is performed by the direct numerical integration in four-dimension. The integration over unit vectors are expressed by that over spherical angles ($\varphi_p,\theta_p,\varphi_q,\theta_q$). Each axis of the space ($\varphi_p,\theta_p,\varphi_q,\theta_q$) is uniformly discretized into $N_{\rm grid}$ points. The integration is then approximated by a summation over the function values at all the grid points in the discretized 4D space:
\begin{equation}
\int {\rm d}\bm{p}\int {\rm d}\bm{q} \ {\rm INT} = \int_0^{2\pi} {\rm d}\varphi_p\int_0^{\pi} {\rm d}\theta_p \int_0^{2\pi} {\rm d}\varphi_q\int_0^{\pi} {\rm d}\theta_q \sin\theta_p \sin\theta_q \cdot {\rm INT} \approx \frac{4\pi^4}{N_{\rm grid}^4} \sum_i \sin\theta_{p,i} \sin\theta_{q,i} \cdot {\rm INT}_{i}
\end{equation}
where $\rm {INT}$ stands for the integrand. We use $N_{\rm grid}=40$ in calculations. This value guarantees that further increasing the $N_{\rm grid}$ will not cause a significant improvement of accuracy while maintaining acceptable computation complexity. Additionally, when evaluating $s_{\rm p^*}$, one-dimensional integrals will be adequate since the integrand is only $\theta_{p^*}$-dependent.

We use a simple iteration approach when evaluating the order parameters $s$ at certain rescaled temperature $k_{\rm B}T/u_{\rm qq}$ via the equation $s=\langle P_2(\cos\theta)\rangle$. Specifically, we start with an initial guess $s_0$, and then $s$ will be updated via a mixing rule: $s_{\rm new} = (1-\epsilon) s_{\rm old} + \epsilon \langle P_2(\cos\theta)\rangle$ for each iteration step where $\epsilon$ is the step size for the update. The iteration proceeds until $|s_{\rm new} - s_{\rm old}| < {\rm tol}$. In our calculation, the update step size is $\epsilon = 0.5$ and tolerance is ${\rm tol} = 10^{-5}$. The iteration stops only if all the order parameters ($s_{\rm p}$, $s_{\rm q}$, and $s_{\rm p^*}$) satisfy the convergence criterion.